\documentclass[prd,onecolumn,preprintnumbers,amsmath,amssymb,nofootinbib,epsfig,bmamsfonts,yfonts,bm]{revtex4}
\usepackage{graphicx}
\usepackage{subfig}

\allowdisplaybreaks

\begin{document}

\title{Kinetic freeze-out, particle spectra and harmonic flow coefficients from mode-by-mode hydrodynamics}

\author{Stefan Floerchinger}
\email{stefan.floerchinger@cern.ch}

\author{Urs Achim Wiedemann}
\email{urs.wiedemann@cern.ch}

\affiliation{Physics Department, Theory Unit, CERN, CH-1211 Gen\`eve 23, Switzerland}

\begin{abstract}
The kinetic freeze-out for the hydrodynamical description of relativistic heavy ion collisions is discussed using a background-fluctuation splitting of the hydrodynamical fields. For a single event, the particle spectrum, or its logarithm, can be written as the sum of background part that is symmetric with respect to azimuthal rotations and longitudinal boosts and a part containing the contribution of fluctuations or deviations from the background. Using a complete orthonormal basis to characterize the initial state allows one to write the double differential harmonic flow coefficients determined by the two-particle correlation method as matrix expressions involving the initial fluid correlations. We discuss the use of these expressions for a mode-by-mode analysis of fluctuating initial conditions in heavy ion
collisions.
\end{abstract}

\maketitle

\section{Introduction}
Fluid dynamic models of relativistic heavy ion collisions at RHIC and at the LHC provide an overall good description of transverse momentum spectra and harmonic flow coefficients up to about $p_T \simeq 2 - 3$ GeV~\cite{Holopainen:2010gz,Qiu:2011hf,Qiu:2011iv,Petersen:2012qc,Qian:2013nba,Gardim:2011xv,Bhalerao:2011bp,Deng:2011at,Niemi:2012aj,Chaudhuri:2012mr,Gale:2012rq,Schenke:2011bn,Rybczynski:2012ed}, for reviews see Refs.~\cite{Heinz:2013th,Gale:2013da,Bozek:2012si}. The initialization of these models needs to account for event wise fluctuations~\cite{Alver:2008zza,Mishra:2007tw,Broniowski:2007ft,Sorensen:2008zk,Takahashi:2009na,Alver:2010gr}, as they may arise either from quantum fluctuations or from the substructure of nuclei in terms of nucleons. Recently, we have developed an 
approach~\cite{Floerchinger:2013rya} that allows for a differential study of the effects of such fluctuations in the fluid dynamic fields. It is based on a mode-by-mode decomposition of fluctuations in single events, on a functional characterization of initial conditions for event classes and on a background-fluctuation splitting for the hydrodynamical evolution. 
Here, we give a technically complete documentation of how kinetic freeze-out of fluctuating modes is handled in this mode-by-mode hydrodynamics. In particular, we 
provide explicit expressions for the hadronic response to specific fluid dynamic fluctuations. These expressions are of more general interest since they can be used to establish to what extent specific fluid dynamic features are washed out or survive hadronization.  

In general, a fluid dynamic description of an expanding system seizes to be valid at kinetic freeze-out, when densities become too low and microscopic interaction times become too long to maintain the system close to local equilibrium. This transition from fluid dynamic behavior to free-streaming particles occurs 
within a finite time after the collision and within finite volume around the collision point. The Cooper-Frye freeze-out prescription~\cite{Cooper:1974mv} is based on the assumptions that freeze-out occurs sufficiently rapidly to
approximate it along a sharp three-dimensional freeze-out hyper surface $\Sigma_f$,
and that the occupation numbers on $\Sigma_f$ are given by free thermal single particle distribution functions~\footnote{Historically, the Cooper-Frye condition was proposed as an improvement of earlier freeze-out conditions going back to the work of Landau~\cite{Landau:1953gs},  and there is some debate on how freeze-out long a sharp hyper surface may be improved further~\cite{Ivanov:2008zi,Bugaev:1999uy}. }. Both assumptions may be questioned. In particular, one often employs in phenomenological applications a smoother freeze-out prescription in which the switch to particle distribution functions at fixed $\Sigma_f$ is followed by an intermediate regime of hadronic scatterings, described for example by Boltzmann's equation, see for example \cite{Pratt:2010jt} for a discussion how this can be done within viscous hydrodynamics. This procedure is interesting since it can give insight into the relevance of a four-dimensional freeze-out volume and a regime of hadronic scattering. However, the usual choice for initializing this regime is again some Cooper-Frye freeze-out prescription, although it would be more consistent to base the initialization in this case on occupation numbers for interacting particles. At least in this sense, the currently used refinements of the Cooper-Frye freeze-out prescription involve additional model assumptions. In the present work, we shall study freeze-out of fluid dynamic fluctuations into free streaming particles without taking hadronic scattering into account, but we shall discuss finally how these effects could be taken into account and where they may be relevant.  

This paper is organized as follows: In section~\ref{sec2}, we first review the Cooper-Frye freeze-out condition for a fluid with finite bulk and shear viscosity. We focus in particular on azimuthally symmetric freeze-out conditions, but we emphasize already here that such  highly symmetric freeze-out conditions are also relevant for collisions at finite impact parameter. The reason is that mode-by-mode hydrodynamics employs a background-fluctuation splitting of all fluid dynamic fields. The background part is chosen to be invariant under rotations in the transverse plane and longitudinal boosts, and the freeze-out hyper surface will inherit these symmetries since we take it as the surface of constant {\it background} temperature. This is slightly different from standard Cooper-Frye prescriptions where $\Sigma_f$ is defined as the surface of constant temperature, but the two descriptions can be mapped to each other to a given order. At linear order in perturbations of the fluid dynamic fields around the background they are actually identical as we show in appendix~\ref{sec:FreezeOutTemperature}. 
All azimuthal asymmetries, irrespective of whether they arise from event-wise fluctuations or from collisions at finite impact parameter, are then encoded in fluctuating modes that are frozen out on the azimuthally symmetric hypersurface $\Sigma_f$. We shall discuss this procedure for the background fields  and for the fluctuating fields in section~\ref{sec3}. 
In section~\ref{sec4} we finally provide explicit expressions for a mode-by-mode decomposition of the fluctuating fields before shortly discussing our results in an outlook and conclusions section. Appendix \ref{appA} compiles some analytic expressions for rapidity and azimuthal integrals and we discuss the difference between a freeze-out at constant background temperature and one at constant total temperature in appendix \ref{sec:FreezeOutTemperature}.

\section{Cooper-Frye kinetic freeze-out and occupation number}
\label{sec2}
Quite generally we assume that at a space-time point $x$ where the system is still close to local thermal equilibrium the occupation number for a specific particle species $i$ takes the form
\begin{equation}
\frac{dN_i}{d^3p d^3x} = f_i(p^\mu; T(x), u^\mu(x), \pi^{\mu\nu}(x), \pi_\text{bulk}(x))\, .
\label{eq:distributionfunct}
\end{equation}
Here, the single particle distribution function $f_i$ depends on the four-momentum
$p^{\mu}$ of the particle, and it depends on the position $x$ through the values that the fluid dynamic fields take at $x$. In equation (\ref{eq:distributionfunct}), we have written the fluid dynamic fields of temperature $T$, velocity $u^{\mu}$, shear stress $\pi^{\mu\nu}$ and bulk viscous pressure $\pi_\text{bulk}$ explicitly. Equivalently, we could have used reparametrizations of these fields, such as  energy density, entropy or enthalpy instead of $T$. And if further fluid dynamic fields would be relevant for characterizing the fluid, the distribution function $f_i$ would depend on them as well. In particular, we neglect here and in what follows non-zero conserved baryon number or electric charge as well as electroagnetic fields, and we do not consider the dependence of occupation numbers on spin. The expression in \eqref{eq:distributionfunct} can be used both for (a set of) on-shell particles where
$E=\sqrt{m_i^2+\vec p^2}$, and for resonances with non-zero width. In that case, $E=\sqrt{\nu_i+\vec p^2}$ and the values of $\nu_i$ are determined by the spectral density $\rho_i(\nu_i)$.

Following the prescription of Cooper and Frye the particle spectrum after kinetic freeze-out is given as
\begin{equation}
E\frac{dN_i}{d^3 p} = -\frac{1}{(2\pi)^3}\,p_\mu \int_{\Sigma_f} d\Sigma^\mu\, f_i(p^\mu;T(x), u^\mu(x), \pi^{\mu\nu}(x), \pi_\text{bulk}(x))\, .
\label{eq:CooperFrye}
\end{equation}
The integral is here over a three-dimensional hyper-surface $\Sigma_f$ where kinetic equilibrium is just still valid. This is sometimes refered to as the "surface of last scattering". The minus sign accounts for our choice of the metric with signature $(-,+,+,+)$.

There is no precise theory on how the freeze-out surface is determined by physical principles.~\footnote{Here, we discuss only kinetic freeze-out. For the chemical freeze-out it has been argued that the QCD phase transition or rapid crossover with decreasing temperature at small baryon densities provides a distinguished event in the typical evolution of a heavy ion collision where the scattering rates drop quickly \cite{BraunMunzinger:2003zz}, leading to $T_\text{fo}\approx T_c$, see also \cite{Heinz:2007in} for arguments in this direction. At high baryon number, or lower collision energy, the chemical freeze-out does not seem to be related to any phase transition but rather too a fixed value for baryon number density \cite{Floerchinger:2012xd}.} 
 The general idea is that it corresponds to the point in the evolution of a fluid element where the scattering rate between particles becomes too small to maintain kinetic equilibrium. This is characterized often by the condition that particle densities become smaller than a certain value corresponding to some freeze-out temperature $T_\text{fo}$.
However, a dynamical freeze-out criterion may be more physical, see for example~\cite{Holopainen:2013jna} for a recent attempt in this direction. This is so since freeze-out should depend also on the history of the collision as can be understood when thinking about a hypothetical very slow expansion where thermal equilibrium would be maintained down to very small temperatures. 

\subsection{Freeze-out at constant background temperature}

As recalled in section~\ref{sec4}, mode-by-mode hydrodynamics is based on a background fluctuation splitting of all fluid dynamic fields. Instead of the standard condition of freeze-out on a hypersurface of constant temperature, we shall work in the following with the related but slightly different prescription of using a hyper surface of constant \emph{background} temperature. For a particular event the actual temperature as determined by the sum of background and fluctuating hydrodynamical fields can then vary on this surface, and the occupation numbers must be corrected accordingly. To any given order in fluctuations, a freeze-out at strictly constant temperature  can be mapped to this description. In principle, one has to include correction terms that account for the change in particle spectra between the two surfaces. To linear order in fluid dynamic fields these correction terms vanish as we show in appendix \ref{sec:FreezeOutTemperature}. In any case, in the absence of a precise theory of freeze-out we do not see a physics argument that would prefer one of these closely related freeze-out conditions.

\begin{figure}
\centering
\includegraphics[width=0.30\linewidth]{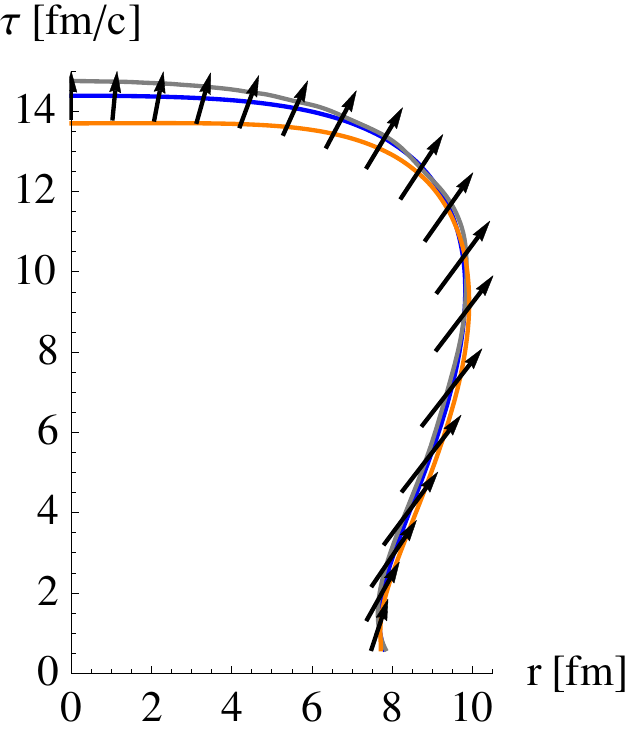}
\caption{Freeze out curve at constant background temperature $T_0=120 \,\text{MeV}$ in the plane of Bjorken time $\tau=\sqrt{t^2-z^2}$ and radius $r=\sqrt{x_1^2+x_2^2}$ for a central Pb-Pb collision at LHC energy. We compare different choices for the shear viscosity to entropy ratio: $\eta/s=0$ (gray, uppermost curve at small $r$), $\eta/s=0.08$ (blue, middle curve at small $r$) and $\eta/s=0.3$ (orange, lowermost curve at small $r$). The arrows indicate the direction of the fluid velocity at freeze out for the case $\eta/s=0.08$.}
\label{fig:FreezeOutCurve}
\end{figure}

In the following, we choose to absorb the entire non-trivial dependence of fluid dynamic fields on azimuthal angle and space-time rapidity in the fluctuating part of fluid dynamic fields. The background part of the fluid dynamic fields is then invariant under azimuthal rotations and Bjorken boost transformations. In general, the freeze out surface at constant temperature will depend on the azimuthal angle $0\leq \varphi < 2\pi$, the space rapidity $-\infty < \eta < \infty $, the proper time $\tau$ and the radius $r$. However, for the symmetric background field considered here, freeze-out on a hyper surface of constant background temperature is characterized for all $\varphi$ and $\eta$ by the same one-parameter curve in the $\tau$-$r$-plane, 
\begin{equation}
\tau=\tau(\alpha), \quad r=r(\alpha)\, ,
\label{eq2.3}
\end{equation}
where without loss of generality we take $\alpha \in [0;1]$. Fig.\ \ref{fig:FreezeOutCurve} shows
the freeze-out curve in the $\tau$-$r$-plane at constant background 
temperature $T_0$ for a central Pb-Pb collision at LHC energies. 
The location of the freeze-out surface, and 
the values of all fluid dynamic fields along it, vary with the value of the 
shear viscosity over entropy ratio, see Fig.\ \ref{fig:FreezeOutCurve}.

%%%%%%%%%%%%%%%%%%%%%%%%%%%%%%%%%%%%%%%
%%%%%%%%%%%%%%%%%%%%%%%%%%%%%%%%%%%%%%%

Since we expand also events at finite impact parameter in terms 
of fluctuations on top of an azimuthally symmetric background, the 
freeze-out for central and non-central Pb-Pb collisions is always 
described by a one-parameter curve of the form (\ref{eq2.3}). 
In practice, one can define this background field for instance as the azimuthally
averaged event sample of the corresponding centrality class~\cite{Floerchinger:2013vua}. 
To formulate freeze-out along (\ref{eq2.3}), we use standard Cartesian 
laboratory coordinates  $t, x_1, x_2, z$ 
\begin{equation}
\begin{split}
& t = \tau(\alpha) \cosh( \eta), \quad x_1 = r(\alpha) \cos (\varphi),\\
& x_2 = r(\alpha) \sin (\varphi), \quad z = \tau(\alpha) \sinh(\eta)\, ,
\end{split}
\end{equation}
and we write the surface element as
\begin{equation}
d\Sigma^\mu =  \tau(\alpha) r(\alpha) \left( \frac{\partial r}{\partial \alpha} \cosh (\eta), \frac{\partial \tau}{\partial \alpha} \cos (\varphi), \frac{\partial \tau}{\partial \alpha} \sin (\varphi), \frac{\partial r}{\partial \alpha} \sinh (\eta) \right) d\alpha d\varphi d\eta.
\end{equation}
For the particle four-momentum 
\begin{equation}
p_\mu=(-E,p_1,p_2,p_3)=\left(-m_T \cosh(\eta_P), p_T \cos(\phi), p_T \sin(\phi), m_T \sinh(\eta_P)\right)\, ,
\end{equation}
we use a parameterization in terms of the transverse mass $m_T^2=E^2-p_3^2$, the transverse momentum $p_T=\sqrt{p_1^2+p_2^2}$, the momentum-space rapidity 
$\eta_P=\text{arctanh}(p_3/E)$ and the (momentum) azimuthal angle 
$\phi=\arctan(p_2/p_1)$. Transverse mass and momentum are related by $m_T^2=m_i^2+p_T^2$ or $m_T^2=\nu_i + p_T^2$ for particles and resonances, respectively. In these coordinates, the freeze-out surface element that enters the 
Cooper-Frye formula (\ref{eq:CooperFrye}) reads
\begin{equation}
-p_\mu d\Sigma^\mu  = \tau(\alpha) r(\alpha) \left( \frac{\partial r}{\partial \alpha} m_T \cosh(\eta_P-\eta) - \frac{\partial \tau}{\partial \alpha} p_T \cos(\phi-\varphi) \right) d\alpha d\varphi d \eta.
\end{equation}
\subsection{Distribution functions at freeze-out}
We now discuss the distribution function $f_i$ in Eq.È \eqref{eq:distributionfunct}. Close to thermal equilibrium, it is of the form
\begin{equation}
f_i=f_{i,\text{eq}}+\delta f_i\, ,
\label{eq:feqdeltafsplit}
\end{equation}
where $f_{i,\text{eq}}$ accounts for the equilibrated part and $\delta f_i$ describes some deviation. The equilibrated part does not depend on the shear stress and the bulk viscous pressure. It is a Lorenz scalar and must therefore be of the form
$f_{i,\text{eq}}=f_{i,\text{eq}}(p_\mu u^\mu, T)$. If kinetic freeze-out happens in a regime where chemical equilibrium is no longer maintained, then the equilibrated occupation numbers $f_{i,\text{eq}}$ also depend on chemical potentials $\mu_i(T)$. In agreement with our approximation to neglect all conserved currents apart from energy and momentum these chemical potentials are not independent in the thermodynamic sense but are functions of the temperature.

If deviations $\delta f_i$ from equilibrium are sufficiently small, then they will be
linear in the shear stress tensor and the bulk viscous pressure. Using that
$\delta f_i$ is a Lorentz scalar and that $u_\mu \pi^{\mu\nu}=0$, one can write
\begin{equation}
\delta f_i = p_\mu p_\nu \pi^{\mu\nu}\;\frac{g_i(p_\mu u^\mu, T,\mu_i)}{(\epsilon + p) T^2}\,  + p_\mu p_\nu \Delta^{\mu\nu}\, \pi_\text{bulk} \, \frac{h_i(p_\mu u^\mu, T,\mu_i)}{(\epsilon+p)T^2},
\label{eq:deltafgh}
\end{equation}
where $\Delta^{\mu\nu} = g^{\mu\nu} + u^\mu u^\nu$. 
The ansatz chosen here makes the functions $g_i$ and $h_i$ dimensionless.
According to this ansatz, the most general distribution function $f_i$ in Eq.\ \eqref{eq:distributionfunct} can be written in terms of the three functions
$f_{i,\text{eq}}$,  $g_{i}$ and $h_{i}$. The explicit form of the distribution 
$f_{i,\text{eq}}$ is of course fixed from the condition of thermodynamic equilibrium. 
However, the functions $g_{i}$ and $h_{i}$ that characterize deviations will depend
on the microscopic dynamics underlying equilibration processes. In principle, these
functions can be determined from a proper, microscopic, quantum field theoretic 
calculation in thermal equilibrium using linear response theory. 

So far we made no assumptions about the size of interaction effects and the discussion was actually not constrained to the regime where perturbation theory or kinetic theory are valid. In particular one may use the formula derived so far to switch from the hydrodynamic description to a more microscopic scattering theory when a particular hyper surface, for example at a certain temperature, is crossed. In general, the equilibrated occupation numbers $f_{i,\text{eq}}$ on such a hyper surface would be those of an interacting system and would differ from the ideal gas form. 
We now leave these general considerations aside and specialize to a freeze-out happening at a point where interaction effects cease to be important. Then, the equilibrated occupation numbers are given by their ideal gas form,
\begin{equation}
f_{i,\text{eq}}(p_\nu u^\nu, T,\mu_i) = \frac{1}{e^\frac{-p_\nu u^\nu-\mu_i}{T} \mp 1}\, ,
\label{eq:freeoccupationnumbers}
\end{equation}
for bosons and fermions respectively. Also, the energy-momentum tensor at a point $x$ is then linear in the occupation numbers for different particle species $i$,
\begin{equation}
T^{\mu\nu} = \sum_i T_i^{\mu\nu} = \int \frac{d^3 p}{(2\pi)^3 E} \, p^\mu p^\nu \, f_i\, .
\label{tmunu}
\end{equation}
Identifying the equilibrated parts on both sides of this equation, one finds for the energy
density and pressure
\begin{equation}
\begin{split}
& \epsilon_i = \int \frac{d^3 p}{(2\pi)^3 E} \, E^2\, f_{i,\text{eq}},\\
& p_i = \int \frac{d^3 p}{(2\pi)^3 E} \, \frac{1}{3} \vec p^2 \, f_{i,\text{eq}}.
\end{split}
\end{equation}
For simplicity we have chosen here a reference frame where $u^\mu=(1,0,0,0)$. Similarly, projecting the non-equilibrated part of (\ref{tmunu}) to the transverse and traceless component yields
\begin{equation}
\pi_i^{jk} = \frac{\pi^{jk}}{(\epsilon+p)T^2} \int \frac{d^3p}{(2\pi)^3 E} \, \frac{2}{15}\, |\vec p|^4 \, g_i\, ,
\label{eq:constraintpi}
\end{equation}
and taking the trace one gets
\begin{equation}
\pi_{\text{bulk},i} = \frac{\pi_\text{bulk}}{(\epsilon+p)T^2}\int \frac{d^3 p}{(2\pi)^3 E} \, \frac{1}{3} |\vec p|^4 \, h_{i}.
\label{eq:constrainth1}
\end{equation}

The form of $\delta f$ in Eq.\ \eqref{eq:deltafgh} makes sure that energy density and pressure are not corrected by the shear viscous part $\sim g_i$. For $h_i$ one has one additional constraint, 
\begin{equation}
\int \frac{d^3 p}{(2\pi)^3 E} E^2 |\vec p|^2\; h_i =0\, ,
\end{equation}
which ensures that energy density is not modified. 
Comparison to \eqref{eq:constrainth1} shows that $h_i=0$ for massless particles.

As mentioned above, the specific form of $g_i$ will depend on the microscopic
dynamics of equilibration processes. To what extent this form is model-dependent
has been discussed for instance in Ref.~\cite{Dusling:2009df}. In what follows,
we shall use the so-called quadratic ansatz  \cite{Teaney:2003kp,Dusling:2009df} 
\begin{equation}
g_i = \frac{1}{2} f_{i,\text{eq}} (1\pm f_{i,\text{eq}}).
\label{eq:gkinetictheory}
\end{equation}
For a single component gas one can check that with this choice \eqref{eq:constraintpi} is fulfilled within a few percent for bosons, fermions and Boltzmann distributed particles. 
Technically, it is convenient to expand $f_{i,\text{eq}}$ in powers of the Boltzmann factor
$\exp\left[\frac{p_\nu u^\nu + \mu_i}{T}\right]$.
 We will therefore work in the following with the occupation numbers
\begin{equation}
f_i = \sum_{j=0}^\infty \left[ 1+\frac{p_\mu p_\nu \pi^{\mu\nu}}{2(\epsilon+p) T^2}(1+j) \right] (\pm1)^j \;e^{\frac{p_\nu u^\nu + \mu_i}{T}(1+j)}
\label{eq:occupationnumbersexpansion}
\end{equation}
for bosons and fermions, respectively.

\section{Particle spectra and background-fluctuation splitting}
\label{sec3}

In section~\ref{sec2}, we have shown that the particle spectrum (\ref{eq:CooperFrye})
can be written as 
\begin{eqnarray}
E\frac{dN_i}{d^3 p} &=& \frac{1}{(2\pi)^3} \int_{\Sigma_f} d\alpha d\varphi d \eta \; 
\tau(\alpha) \, r(\alpha) \left( \frac{\partial r}{\partial \alpha} m_T \cosh(\eta_P-\eta) - \frac{\partial \tau}{\partial \alpha} p_T \cos(\phi-\varphi) \right) \nonumber \\
&& \times
\sum_{j=0}^\infty \left[ 1+\frac{p_\mu p_\nu \pi^{\mu\nu}}{2(\epsilon+p) T^2}(1+j) \right] (\pm1)^j \;e^{\frac{p_\nu u^\nu + \mu_i}{T}(1+j)}\, ,
 \label{eq3.1}
\end{eqnarray}
where integration is over an azimuthally symmetric and Bjorken boost invariant 
hyper surface $\Sigma_f$. The fluid dynamic fields $u_\mu$, $T$, $\pi^{\mu\nu}$
and $\mu_i$ that appear in (\ref{eq3.1}) are evaluated on $\Sigma_f$. 
Choosing this freeze-out to occur at constant background temperature
$T_0$ means that we perform a background-fluctuation splitting of the 
temperature into a position-independent background and a position-dependent
fluctuation term,
\begin{equation}
	T(\alpha,\varphi,\eta) = T_0 + \delta T(\alpha,\varphi,\eta)\, .
	\label{eq3.2}
\end{equation}
For all other fluid dynamic fields, the background
terms can depend on the position along $\Sigma_f$, but they share the 
symmetries of $\Sigma_f$. We therefore write for the background-fluctuation splitting 
\begin{eqnarray}
	u^{\mu}(\alpha,\varphi,\eta) &=& u^{\mu}_0(\alpha) 
	  + \delta u^{\mu}(\alpha,\varphi,\eta)\, ,
	  \label{eq3.3} \\
	  \pi^{\mu\nu}(\alpha,\varphi,\eta) &=& \pi^{\mu\nu}_0(\alpha) 
	  + \delta \pi^{\mu\nu}(\alpha,\varphi,\eta)\, ,
	  \label{eq3.4}
\end{eqnarray}
and likewise for other fluid dynamic fields such as $\pi_\text{Bulk}$. The chemical potentials $\mu_i$ are assumed to depend only on temperature $T$ so that they can be written as in eq.\ (\ref{eq3.2}). We recall that the 
ansatz (\ref{eq3.2})-(\ref{eq3.4}) is valid also for collisions at finite impact
parameter, since it allows for the parametrization of arbitrary $\varphi$-dependencies 
of the fluid fields. One way to see that the background-fluctuation splitting (\ref{eq3.2})-(\ref{eq3.4}) is physically meaningful is to consider the background parts of all fields
as characterizing the event average over many collisions (with random azimuthal 
orientation), while the fluctuating parts characterize event-specific fluctuations. 
For more technical details on this point, see Ref.~\cite{Floerchinger:2013vua}.

The strategy of the following is to use the ansatz (\ref{eq3.2})-(\ref{eq3.4}) to expand the spectrum (\ref{eq3.1}) in fluctuations around the background fields. We do this by 
evaluating first the zeroth order background contribution to (\ref{eq3.1}) in 
subsection~\ref{sec3a}. We then discuss in subsection ~\ref{sec3b} the contributions
that arise from first order in fluctuations. We anticipate that the final result of this 
excercise will be a set of equations that relate in a very explicit form specific modes 
of fluctuating hydrodynamic fields to specific features in experimental observables. 

%%%%%%%%%%%%%%%%%%%%%%%%%%%%%%%%%%%%%%%%%%
\subsection{Zeroth order in fluctuations: the background contribution} 
 \label{sec3a}

To zeroth order in fluctuations, the temperature-dependence of the spectrum (\ref{eq3.1})
reduces to a dependence on $T_0$. As for the background fluid velocity $u_0^{\mu}$, 
symmetries imply that only the temporal and radial component can take non-vanishing values
in Bjorken coordinates $\tau, r, \varphi, \eta$. The Lorentz scalar $p_\mu u_0^\mu$ in  (\ref{eq3.1}) hence takes the form
\begin{equation}
p_\mu u_0^\mu = -m_T \,u_0^\tau\, \cosh(\eta_P-\eta) + p_T\, u_0^r\, 
\cos(\phi-\varphi)\, .
\end{equation}
For the background part of the shear stress tensor symmetries imply that only the components $\pi_0^{\tau\tau}$, $\pi_0^{\tau r}$, $\pi_0^{r \tau}$, $\pi_0^{r r}$, $\pi_0^{\varphi \varphi}$ and $\pi_0^{\eta\eta}$ are non-zero. Due to the transverse and traceless constraints only two of them are actually independent. We choose a parameterization of the form (still in coordinates $(\tau,r,\varphi,\eta)$)
\begin{equation}
\pi_0^{\mu\nu} = (\epsilon_0+\pi_0) 
\begin{pmatrix}
(u_0^r)^2 (\tilde \pi_0^t-\tfrac{1}{2}\tilde \pi_0^{\eta\eta}) && u_0^\tau u_0^r (\tilde \pi_0^t-\tfrac{1}{2}\tilde \pi_0^{\eta\eta}) && 0 && 0 \\
u_0^\tau u_0^r (\tilde \pi_0^t-\tfrac{1}{2}\tilde \pi_0^{\eta\eta}) && (u_0^\tau)^2  (\tilde \pi_0^t-\tfrac{1}{2}\tilde \pi_0^{\eta\eta}) && 0 && 0 \\
0 && 0 && \frac{1}{r^2}(-\tilde \pi_0^t-\frac{1}{2}\tilde \pi_0^{\eta\eta}) && 0 \\
0 && 0 && 0 && \frac{1}{\tau^2} \pi_0^{\eta\eta}
\label{eq3.6}
\end{pmatrix}.
\end{equation}
Here and in the following, we denote by a tilde suitably rescaled field components.
Entering with these equations the single inclusive hadron 
spectrum (\ref{eq3.1}) and performing the integrals over space rapidity $\eta$ 
and azimuthal angle $\varphi$ using the identities in appendix \ref{sec:RapidtyAzimuthalIntegrals}, one finds 
\begin{equation}
\begin{split}
S_{0}(m_T,p_T) = & \frac{1}{2\pi^2}\int d\alpha\,\tau(\alpha) r(\alpha) \sum_{j=0}^\infty (\pm 1)^j \;\;e^{\frac{\mu_{i0}}{T_0}(1+j)}\\
\times & {\Bigg [} \frac{d r}{d\alpha} m_T K_1\left( \frac{m_T u_0^\tau}{T_0}(1+j) \right) I_0\left( \frac{p_T u_0^r}{T_0} (1+j) \right) \\
&- \frac{d\tau}{d\alpha} p_T K_0\left( \frac{m_T u_0^\tau}{T_0}(1+j) \right) I_1\left( \frac{p_T u_0^r}{T_0} (1+j) \right)\\
& + \frac{d r}{d\alpha} m_T \; \tilde\pi_0^t \, Z_1\left( \frac{m_T}{T_0}, \frac{p_T}{T_0}, u_0^\tau, u_0^r, j \right) - \frac{d\tau}{d\alpha} p_T \; \tilde\pi_0^t \, Z_2\left( \frac{m_T}{T_0}, \frac{p_T}{T_0}, u_0^\tau, u_0^r, j \right)\\
& + \frac{d r}{d\alpha} m_T \; \tilde\pi_0^{\eta\eta} \, Z_3\left( \frac{m_T}{T_0}, \frac{p_T}{T_0}, u_0^\tau, u_0^r, j \right) - \frac{d\tau}{d\alpha} p_T \; \tilde\pi_0^{\eta\eta} \, Z_4\left( \frac{m_T}{T_0}, \frac{p_T}{T_0}, u_0^\tau, u_0^r, j \right) {\Bigg ]}\, .
\end{split}
\label{eq:S0}
\end{equation}
Here, $I_n(z)$ and $K_n(z)$ are modified Bessel functions of the first and second kind, respectively, and the integration kernels $Z_1$, $Z_2$, $Z_3$ and $Z_4$ are given in the  appendix~\ref{sec:RapidtyAzimuthalIntegrals}.

%%%%%%%%%%%%%%%%%%%%%%%%%%%%%%%
\begin{figure}[h]
\centering
\subfloat[]{\includegraphics[width=0.43\linewidth]{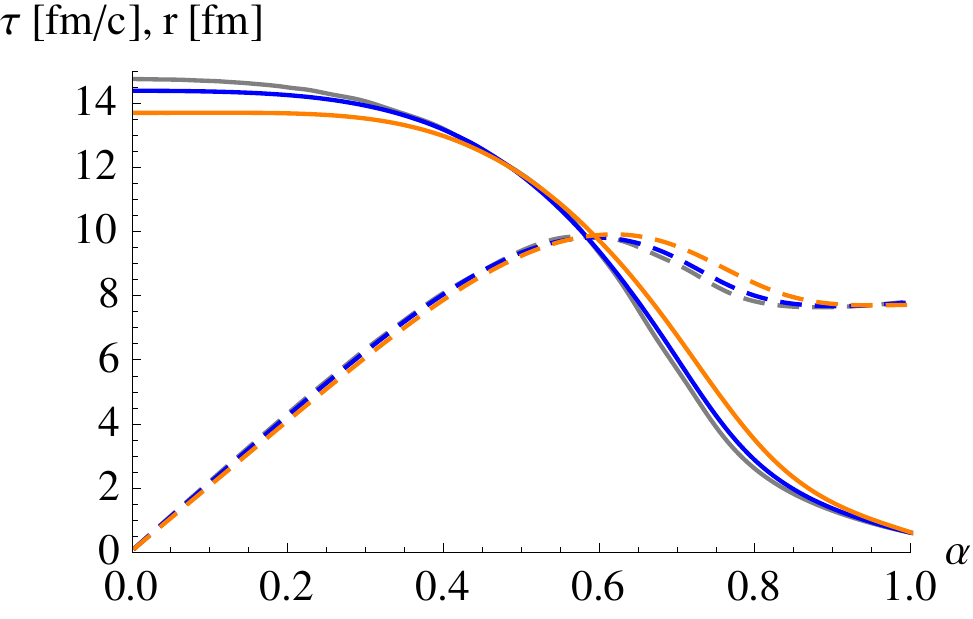}}
\subfloat[]{\includegraphics[width=0.48\linewidth]{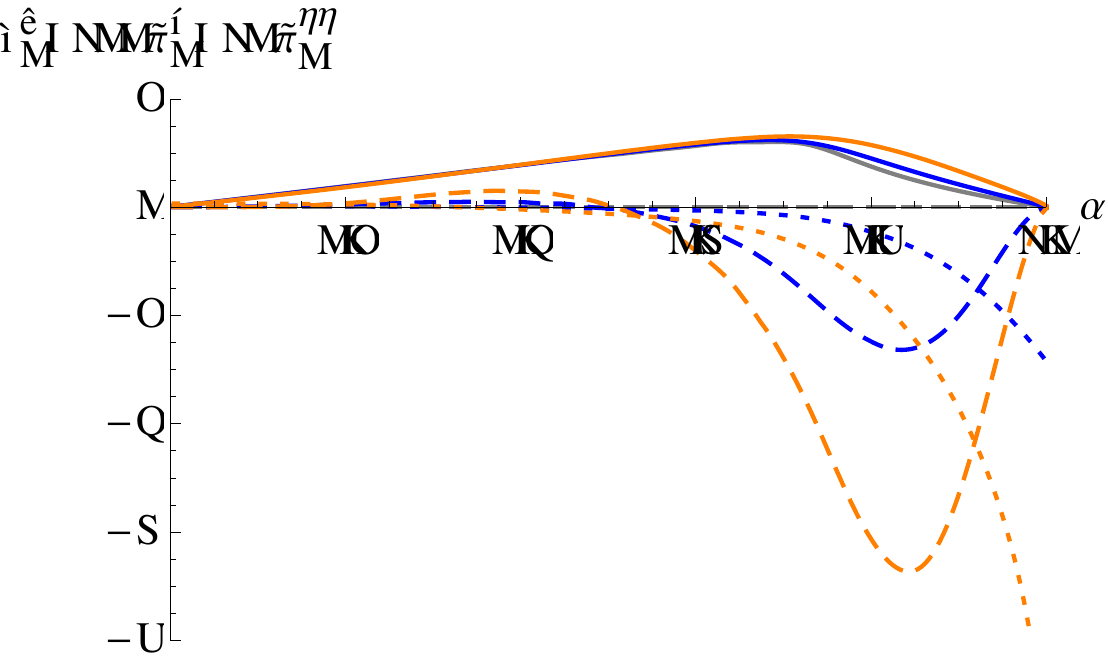}}
\caption{(a) Freeze out time $\tau$ (solid) and radius $r$ (dashed) as a function of the parameter $\alpha$ which parameterizes the freeze-out surface. 
(b) Background fluid velocity $u^r_0$ (solid) and shear stress components $\tilde \pi^t_0$ (dashed) and $\tilde \pi^{\eta\eta}_0$ (dotted) as a function of the parameter $\alpha$. 
We compare different choices for the shear viscosity to entropy ratio with the same color coding as in Fig.\ \ref{fig:FreezeOutCurve}.}
\label{fig:FreezeOutCurve2}
\end{figure}
%%%%%%%%%%%%%%%%%%%%%%%%%%%%%%%%%

For a given form of the independent hydrodynamic fields $T_0$, $u_0^r$, $\tilde \pi^t_0$ and $\tilde \pi_0^{\eta\eta}$, eq.\ \eqref{eq:S0} determines the corresponding background particle spectrum as an integral over the freeze-out hyper surface at fixed background
temperature. It can be evaluated for different particle species by using the appropriate relation between $m_T$ and $p_T$ and adding the appropriate degeneracy factors for spin and isospin. In Fig.\ \ref{fig:FreezeOutCurve2} we show the functions $\tau(\alpha)$, $r(\alpha)$, $u_0^r(\alpha)$, $\tilde \pi_0^t(\alpha)$ and $\tilde \pi_0^{\eta\eta}(\alpha)$ for a hydrodynamic calculation corresponding to central Pb-Pb collisions at the LHC. The parameters such as initialization time, initial temperature etc. have been chosen as in Ref.\ \cite{Shen:2012vn} and the freeze-out curve corresponds to Fig.\ \ref{fig:FreezeOutCurve}.

\begin{figure}
\centering
\subfloat[]{\includegraphics[width=0.48\linewidth]{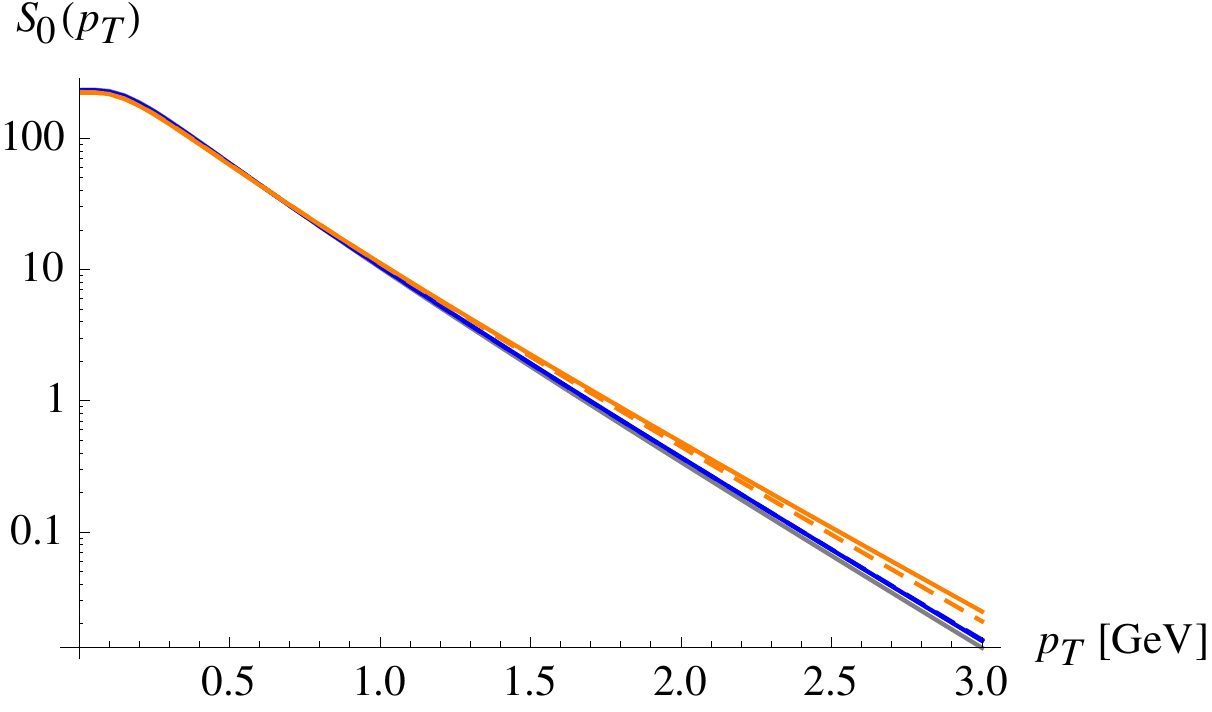}}
\subfloat[]{\includegraphics[width=0.48\linewidth]{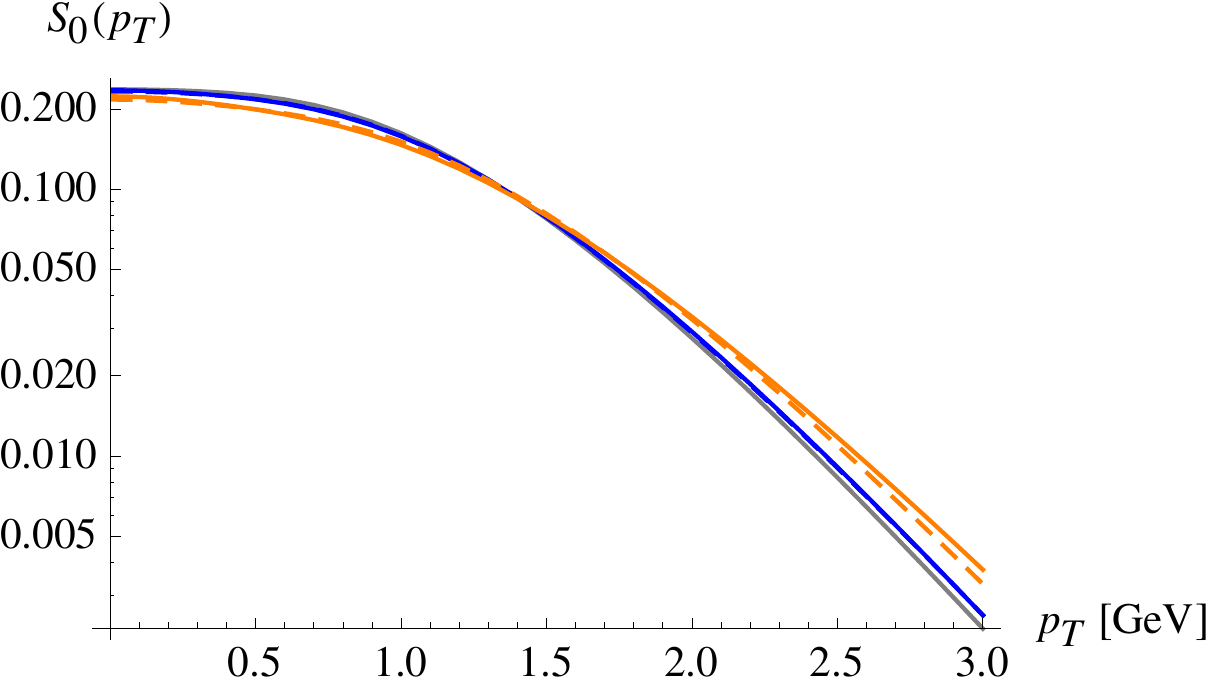}}
\caption{Background contribution to the particle spectrum $S_0(p_T)$ for pions (a) and protons (b). Spin or isospin degeneracy factors are not taken into account. The curves are for different values of the shear viscosity to entropy ratio with the same color coding as in Fig.\ \ref{fig:FreezeOutCurve}. We also show the curves obtained if the shear stress contribution at freeze out (``quadratic ansatz") is neglected (dashed lines).}
\label{fig:particleSpectrumS0}
\end{figure}

In Fig.\ \ref{fig:particleSpectrumS0} we plot the background contribution $S_0(m_T,p_T)$ to the single inclusive transverse momentum spectrum of pions and protons, neglecting spin and iso-spin degeneracy factors as well as chemical potentials.   We compare different values of the shear viscosity to entropy ratio.  We also show the result obtained without the term linear in the shear stress, i.\ e.\ with $\tilde \pi_0^t = \tilde \pi_0^{\eta\eta}=0$ in eq.\ \eqref{eq:S0}. The result changes only slightly: At LHC energies the lifetime of the fireball is long enough for the background part of the shear stress to relax largely to its equilibrium value.

\subsection{Particle spectra and harmonic flow due to fluctuating fields}
\label{sec3b}

We discuss now how fluctuations around the background hydrodynamic fields
contribute to the particle spectrum at freeze-out. To that end
we first introduce a specific parametrization for fluctuations in all scalar, vector and
tensor fields. For the enthalpy density $w=\epsilon + p$ and the fluid velocity on the freeze-out surface, we write to lowest order in the deviations from the background fields
\begin{equation}
\begin{split}
& w=w_0 + w_0 \tilde w\, ,\quad \quad  
u^\tau = u_0^\tau + \frac{u_0^r}{u_0^\tau} \tilde u^r\, ,\\
& u^r = u_0^r + \tilde u^r, \quad \quad u^\varphi = \frac{1}{r} \tilde u^\varphi, \quad \quad u^\eta = \frac{1}{\tau} \tilde u^\eta.
\end{split}
\end{equation}
Here, the fluctuating part is parametrized such that 
$\tilde w$, $\tilde u^r$, $\tilde u^\varphi$ and $\tilde u^\eta$ are dimensionless. 
It is useful to introduce a circular polarization basis for the transverse components 
of the vector field,
\begin{equation}
\tilde u^r = \frac{1}{\sqrt{2}} \left( \tilde u^- + \tilde u^+ \right), \quad\quad \tilde u^\varphi = \frac{1}{\sqrt{2}} \left( i\, \tilde u^- - i\, \tilde u^+ \right).
\end{equation}
For the shear stress tensor, we parametrize fluctuations around the background field 
to linear order in the form
\begin{equation}
\pi^{\mu\nu} = \pi_0^{\mu\nu} + (\epsilon_0+p_0) 
\begin{pmatrix}
\tilde \pi^{\tau\tau} && \tilde \pi^{\tau r} && \frac{1}{r} \tilde \pi^{\tau\varphi} && \frac{1}{\tau} \tilde \pi^{\tau\eta} \\
\tilde \pi^{r\tau} && \tilde \pi^{rr} && \frac{1}{r} \tilde \pi^{r\varphi} && \frac{1}{\tau} \tilde \pi^{r\eta} \\
\frac{1}{r} \tilde \pi^{\varphi\tau} && \frac{1}{r} \tilde \pi^{\varphi r} && \frac{1}{r^2} \tilde \pi^{\varphi\varphi} && \frac{1}{\tau r} \tilde \pi^{\varphi\eta} \\
\frac{1}{\tau} \tilde \pi^{\eta\tau} && \frac{1}{\tau} \tilde \pi^{\eta r} && \frac{1}{\tau r} \tilde \pi^{\eta \varphi} && \frac{1}{\tau^2} \tilde \pi^{\eta\eta}
\end{pmatrix}.
\end{equation}
Not all components of this shear stress tensor are independent. There is a constraint from
the symmetry $\pi^{\mu\nu}=\pi^{\nu\mu}$ that implies 
$\tilde\pi^{\mu\nu}=\tilde \pi^{\nu\mu}$. Also, the shear stress tensor is traceless, 
$\pi^\mu_\mu=0$ and hence 
$\tilde \pi^{\tau\tau} = \tilde \pi^{rr} + \tilde \pi^{\varphi\varphi} + \tilde \pi^{\eta\eta}$.
In addition, there is the orthogonality relation $u_\mu \pi^{\mu\nu}=0$ that implies
to linear order in the fluctuation field
$(u_\mu-(u_0)_\mu) \pi_0^{\mu\nu} + (u_0)_\mu (\pi^{\mu\nu}-\pi_0^{\mu\nu}) = 0$.
As a consequence of these constraints, 
only five components of $\tilde\pi^{\mu\nu}$ are independent. We choose them to be $\tilde \pi^{r\varphi}$, $\tilde \pi^{\varphi\varphi}$, $\tilde \pi^{r\eta}$, $\tilde\pi^{\varphi\eta}$ and $\tilde \pi^{\eta\eta}$. Again it is useful to employ a circular polarization basis writing
\begin{equation}
\begin{split}
\tilde \pi^{r\varphi} = \frac{1}{\sqrt{2}} \left( \tilde \pi^{--} + \tilde \pi^{++} \right), \quad & \quad \tilde \pi^{\varphi\varphi} = \frac{1}{\sqrt{2}} \left( i \tilde \pi^{--} - i \tilde \pi^{++} \right)-\frac{1}{2}\tilde \pi^{\eta\eta},\\
\tilde \pi^{r\eta} = \frac{1}{\sqrt{2}} \left( \tilde \pi^{-\eta} + \tilde \pi^{+\eta} \right), \quad & \quad \tilde \pi^{\varphi\eta} = \frac{1}{\sqrt{2}} \left( i\tilde \pi^{-\eta} - i \tilde \pi^{+\eta} \right).
\end{split}
\end{equation}
The dependent components of the shear stress tensor can be expressed in terms of the independent ones as well as the velocity field,

\begin{equation}
\begin{split}
& \tilde \pi^{\tau\tau} = 2 u_0^r \left( \tilde \pi_0^t - \tfrac{1}{2}\tilde \pi_0^{\eta\eta} \right) \tfrac{1}{\sqrt{2}} \left(\tilde u^-+\tilde u^+\right) - (u_0^r)^2 \tfrac{1}{\sqrt{2}} \left( i \tilde \pi^{--} - i \tilde \pi^{++} \right) - \tfrac{1}{2}(u_0^r)^2 \tilde \pi^{\eta\eta},\\
& \tilde \pi^{\tau r} = \left( u_0^\tau + \tfrac{(u_0^r)^2}{u_0^\tau} \right) \left( \tilde \pi_0^t - \tfrac{1}{2}\tilde \pi_0^{\eta\eta} \right)\tfrac{1}{\sqrt{2}} \left(\tilde u^-+\tilde u^+\right) - u_0^\tau u_0^r \tfrac{1}{\sqrt{2}} \left( i \tilde \pi^{--} - i \tilde \pi^{++} \right) - \tfrac{1}{2} u_0^\tau u_0^r \tilde \pi^{\eta\eta},\\
& \tilde \pi^{r r} = 2 u_0^r \left( \tilde \pi_0^t - \tfrac{1}{2}\tilde \pi_0^{\eta\eta} \right) \tfrac{1}{\sqrt{2}} \left(\tilde u^-+\tilde u^+\right) - (u_0^\tau)^2 \tfrac{1}{\sqrt{2}} \left( i \tilde \pi^{--} - i \tilde \pi^{++} \right) - \tfrac{1}{2} (u_0^\tau)^2 \tilde \pi^{\eta\eta},\\
& \tilde \pi^{\tau \varphi} = -\frac{1}{ u_0^\tau} \left( \tilde \pi_0^t + \tfrac{1}{2}\tilde \pi_0^{\eta\eta} \right) \tfrac{1}{\sqrt{2}} \left(i \tilde u^- - i \tilde u^+\right) + \tfrac{u_0^r}{u_0^\tau} \tfrac{1}{\sqrt{2}} \left( \tilde \pi^{--} + \tilde \pi^{++} \right),\\
& \tilde \pi^{\tau \eta} = \frac{1}{ u_0^\tau} \tilde \pi_0^{\eta\eta} \tilde u^\eta + \tfrac{u_0^r}{u_0^\tau} \tfrac{1}{\sqrt{2}} \left( \tilde \pi^{-\eta} + \tilde \pi^{+\eta} \right).
\end{split}
\end{equation}

In our construction the background fields $w_0$, $u_0^\mu$,  $\pi_0^{\mu\nu}$  and the freeze-out surface are symmetric with respect to azimuthal rotations and Bjorken boost transformations. It is therefore useful to make a Fourier expansion for the fluctuation field according to
\begin{equation}
\begin{split}
& \tilde w(\tau(\alpha), r(\alpha), \varphi, \eta)  = \sum_{m=-\infty}^\infty \int \frac{d k_\eta}{2\pi} \; \tilde w(\tau(\alpha), r(\alpha), m ,k_\eta) \; e^{im\varphi + i k_\eta \eta},\\
& \tilde w(\tau(\alpha), r(\alpha), m, k_\eta)^*  = \tilde w(\tau(\alpha), r(\alpha), -m, -k_\eta),
\label{eq:FourierExpFluct}
\end{split}
\end{equation}
and similar for $\tilde u^-$, $\tilde u^+$ and $\tilde u^\eta$ and the fluctuations 
$\tilde\pi^{\mu\nu}$ in the shear viscous tensor. 
On the linear level, one can determine the contribution of each mode to the spectrum separately. For symmetry reasons, this contribution will be proportional to a phase factor
$e^{i m\phi + i k_\eta \eta_P}$, where $\phi$ and $k_\eta$ denote the momentum space azimuthal angle and rapidity, respectively. We introduce the short hand notation
\begin{equation}
\tilde w(\alpha) = \tilde w(\tau(\alpha), r(\alpha), m, k_\eta)\, ,
\end{equation}
etc.  On the freeze-out hyper surface, each mode is then characterized by the set of 
ten functions which are  $\tilde w(\alpha)$, $\tilde u^-(\alpha)$, $\tilde u^+(\alpha)$ 
and $\tilde u^\eta(\alpha)$ and the five independent components of 
$\tilde\pi^{\mu\nu}(\alpha)$. From equation (\ref{eq3.1}), we find for the 
contribution of a each single fluctuating mode to the spectrum up to linear order 
\begin{equation}
\begin{split}
\tilde S(m_T, p_T) = & \frac{1}{(2\pi)^3} \int d\alpha \int_0^{2\pi} d\varphi \int_{-\infty}^\infty d\eta\;\; \tau(\alpha) r(\alpha) \left[ \frac{d r}{d\alpha} m_T \cosh(\eta) - \frac{d\tau}{d\alpha} p_T \cos(\varphi) \right] \\
& \times \sum_{j=0}^\infty (\pm1)^j \;\;\text{exp} \left( -\frac{m_T u_0^\tau}{T_0} (1+j) \cosh(\eta) + \frac{p_T u_0^r}{T_0} (1+j) \cos(\varphi) + \frac{\mu_{i0}}{T_0} \right) (1+j)\\
&\times {\Bigg [} \left( \frac{m_T u_0^\tau}{T_0} \cosh(\eta) - \frac{p_T u_0^r}{T_0}\cos(\varphi) - \frac{\mu_{i0}}{T_0}+\frac{d \mu_{i0}}{dT_0}\right) \frac{d \ln (T_0)}{d \ln (w_0)} \;\; \tilde w(\alpha)\\
& \quad + \left( -\frac{m_T u_0^r}{T_0 u_0^\tau \sqrt{2}} \cosh(\eta) + \frac{p_T}{T_0\sqrt{2}} e^{-i\varphi} \right)\;\; \tilde u^-(\alpha)\\
& \quad + \left( -\frac{m_T u_0^r}{T_0 u_0^\tau \sqrt{2}} \cosh(\eta) + \frac{p_T}{T_0\sqrt{2}} e^{i\varphi} \;\; \right) \;\; \tilde u^+(\alpha) \\
& \quad + \left( -\frac{m_T}{T_0} \sinh(\eta) \right)\;\; \tilde u^\eta(\alpha)  + \hbox{(Terms 
proportional to $\tilde\pi^{\mu\nu}(\alpha)$)} {\Bigg ]}\, e^{im\varphi + i k_\eta \eta}\, .
\end{split}
\label{eq:spectrumfluctbeforeintegration}
\end{equation}
The contributions from fluctuations in the shear viscous tensor are not written out explicitly 
in this intermediary result (\ref{eq:spectrumfluctbeforeintegration}), but we present them
for the final result (\ref{eq:spectrumfluctkernels}) below. To arrive at this final 
result, we integrate analytically over the azimuthal angle $\varphi$ and rapidity $\eta$. 
The final result takes then the form
\begin{equation}
\begin{split}
\tilde S&(m_T,p_T) = \frac{1}{2\pi^2} \int d\alpha \; \tau(\alpha) r(\alpha) \;\sum_{j=0}^\infty (\pm1)^j (1+j)\, \sum_{i=1}^{10}\\
& \times {\Bigg [} 
\frac{dr}{d\alpha}\; m_T\;  \tilde h_i(\alpha)\; Y_{i,a} \left(m,k_\eta;\tfrac{m_T}{T}, \tfrac{p_T}{T},h^{BG}(\alpha),j\right) - \frac{d\tau}{d\alpha} \;p_T\;  \tilde h_i(\alpha)\; Y_{i,b} \left(m,k_\eta;\tfrac{m_T}{T}, \tfrac{p_T}{T}, h^{BG}(\alpha),j  \right)  {\Bigg ]}.
\end{split}\label{eq:spectrumfluctkernels}
\end{equation}
Here and in what follows, we compose the background fields into
the structure $h^{BG}$, with ten independent 
field components
\begin{equation}
h^{BG}= \left(w_0, u_0^r, u_0^\phi, u_0^\eta, \pi_{0\, \text{bulk}},
\pi_0^{--}, \pi_0^{++},\pi_0^{-\eta}, \pi_0^{+\eta}, \pi_0^{\eta\eta} \right)\, . 
\label{eq3.17}
\end{equation}
Analogously, we denote fluctuations around these background fields by the 
shorthand $\tilde h$, 
\begin{equation}
\tilde h= \left(\tilde w, \tilde u^r, \tilde u^\phi, \tilde u^\eta, \tilde \pi_\text{bulk},
\tilde \pi^{--}, \tilde \pi^{++},\tilde \pi^{-\eta}, \tilde \pi^{+\eta}, \tilde \pi^{\eta\eta} \right)\, ,
\label{eq3.18}
\end{equation}
with $i$-th component $\tilde h_i$. 
As seen from the structure of equation (\ref{eq:spectrumfluctkernels}),
the integral kernels $Y_{i,a}$ and $Y_{i,b}$ specify the response of the single inclusive
hadron spectrum to fluctuations on the freeze-out surface in enthalpy, fluid velocity,
bulk viscous pressure and shear stress tensor. These kernels are given explicitly  in appendix~\ref{sec:RapidtyAzimuthalIntegrals}. They depend in general on the value
$h^{BG}(\alpha)$ of the background field along the freeze-out hyper surface. 
They also depend on the summation index $j$ as specified in the appendix \ref{sec:RapidtyAzimuthalIntegrals}.

In summary, the output of any fluid dynamical simulation can be 
written in terms of a set of azimuthally symmetric background fields $h^{BG}(\alpha)$
along the freeze-out hyper surface, supplemented by a set of fluctuations $\tilde h(\alpha)$. 
Equation \eqref{eq:spectrumfluctkernels} specifies the sensitivity of the single inclusive 
hadron spectrum up to linear order in all fluctuations. It can be used to 
describe the particle distribution for a particular event. Also, event averages of
\eqref{eq:spectrumfluctkernels}  can be determined if the 
the probability distribution of the fluctuations $\tilde h$ for an event class is known.

\section{Mode-by-mode decomposition}
\label{sec4}

Mode-by-mode hydrodynamics~\cite{Floerchinger:2013rya} provides a more general strategy for characterizing single fluctuating events, but also event averages, correlations 
and probability distributions. It is based on decomposing arbitrary fluctuating initial conditions
(in all scalar, vector and tensor fields $\tilde h^{(i)}$) in a complete orthonormal set of basis functions. For each of these basis functions ("modes"), the (linearized) hydrodynamic equations can then be solved. Knowing the decomposition of initial conditions in terms of orthonormal modes and the dynamical propagation of each mode, one has then
a direct bridge between initial conditions and particle spectra for single events and event averages. 

Here, we illustrate this strategy first for cases for which the initial conditions do not show
fluctuations in the fluid velocity and shear stress, so that the contribution $\tilde w$ to
the enthalpy density is the only origin of fluctuations. This case is most often assumed in 
the phenomenological literature. For simplicity we also neglect first all dependence on rapidity $\eta$ so that the right hand side of \eqref{eq:FourierExpFluct} has support for $k_\eta=0$ only. Under these assumptions, we discuss the effect of fluctuations on single inclusive
hadron spectra in section~\ref{sec4a}, and on two-particle correlation functions in
section~\ref{sec4b}. In section~\ref{sec4c}, we discuss then shortly how these assumptions
can be relaxed.

\subsection{Single inclusive hadron spectra}
\label{sec4a}
Following Ref.~\cite{Floerchinger:2013vua} 
we decompose the initial fluctuations in an orthonormal basis,
\begin{equation}
\tilde w(\tau_0,r,\varphi) = \sum_{m=-\infty}^{\infty} \sum_{l=1}^\infty \tilde w_l^{(m)}\, e^{i m\varphi} \, f^{(m)}_l\left(r \right),
\label{eq:FourierBesselEnthalpy}
\end{equation}
where $f^{(m)}_l(r)$ is an appropriate set of basis functions.   Evolving 
the set of basis functions $f^{(m)}_l(r)$ in (linearized) hydrodynamics 
from initial time $\tau_0$ onwards gives rise to fluctuations $\tilde h_i$ in all 
fluid dynamic fields at later times. Inserting the fluctuations $\tilde h_i(\alpha)$ 
at freeze-out  thus obtained into equation \eqref{eq:spectrumfluctkernels}, one 
establishes a linear dynamical mapping from modes $f^{(m)}_l(r)$ in the initial
conditions to corresponding contributions $\tilde S^{(m)}_l(m_T,p_T)$ in the 
final hadronic spectrum, 
\begin{equation}
f^{(m)}_l(r) \longrightarrow \tilde S^{(m)}_l(m_T,p_T)\, .
\label{dynamicalmap}
\end{equation}
The complete particle spectrum for the event in \eqref{eq:FourierBesselEnthalpy} is then 
given to first order in the fluctuations by the linear superposition of modes with 
wave numbers $(m, l)$ on top of the background spectrum $S_0(m_T,p_T) $,
\begin{equation}
\left( \frac{dN}{p_T dp_T d\phi d\eta_P} \right)_\text{single event} = S_0(m_T,p_T) + \sum_{m=-\infty}^\infty \sum_{l=1}^\infty \,\tilde w^{(m)}_l \, e^{i m \phi}\, \tilde S^{(m)}_l(m_T,p_T).
\end{equation}
We note that in this expression, events are characterized by the set of weights 
$\tilde w^{(m)}_l$ with which the fluctuating modes $f^{(m)}_l(r)$ are 
present in the initial conditions. Therefore, once one has determined the dynamical
mapping \eqref{dynamicalmap} for the orthonormal basis functions $f^{(m)}_l(r)$,
the hydrodynamic evolution of different events is given without further
fluid dynamic simulation simply by changing the set of weights $\tilde w^{(m)}_l$
in the spectrum above. 

In a recent paper~\cite{Floerchinger:2013vua}, we have shown how the initial 
conditions can be characterized for event ensembles in terms of a probability distributions 
$p\left[ \tilde w\right]$ of the weights $\tilde w^{(m)}_l$ in 
\eqref{eq:FourierBesselEnthalpy}. For the phenomenologically relevant case of 
determining the particle spectrum as an average over many events, one can 
write then up to linear order in the fluctuations 
\begin{equation}
\begin{split}
\left( \frac{dN}{p_T dp_T d\phi d\eta_P} \right)_\text{event average} = & S_0(m_T,p_T) + \int D\tilde w\; p[\tilde w]\;\sum_{m=-\infty}^\infty \sum_{l=1}^\infty \,\tilde w^{(m)}_l \, e^{i m \phi}\, \tilde S^{(m)}_l(m_T,p_T)\\
= &  S_0(m_T,p_T) + \sum_{m=-\infty}^\infty \sum_{l=1}^\infty \langle \tilde w^{(m)}_l \rangle \, e^{i m \phi}\, \tilde S^{(m)}_l(m_T,p_T)\, ,
\end{split}
\label{linearinS}
\end{equation}
where the measure $D\tilde w$ denotes integration over all modes $\tilde w^{(m)}_l$,
and $\langle f \rangle = \int D\tilde w\; p[\tilde w] f $. For an azimuthally
symmetric background, event averages $\langle \tilde w^{(m)}_l \rangle$ vanish
for $m\not= 0$ and fluctuations lead to non-vanishing event averages only for 
correlations amongst the
weights of modes, such as $\langle \tilde w^{(m)}_l\, w^{(m')}_{l'} \rangle$. 
In this case, the particle spectrum 
\eqref{linearinS} corresponds therefore to the background contribution $S_0$ plus
a possible contributions proportional to $\langle \tilde w^{(0)}_l \rangle$. When the background is taken to be the event average of the fluid fields, the expectation values $\langle \tilde w_l^{(0)} \rangle$ vanishes as well.

Instead of expanding the particle spectrum in the small fluctuations as above it may actually be advantageous to expand its logarithm. The underlying observation is that the contribution of a particular fluid cell to the particle spectrum at freeze-out depends on the hydrodynamic fields in an essentially exponential way. We write
\begin{equation}
\ln \left( \frac{dN}{p_T dp_T d\phi d\eta} \right)_\text{single event} = \ln S_0(m_T, p_T) + \sum_{m=-\infty}^\infty \sum_{l=1}^\infty \; \tilde w^{(m)}_l \, e^{im\phi} \; \theta^{(m)}_l(m_T,p_T)\, ,
\end{equation}
where the response of the logarithm of the spectrum to a single basis mode is now given 
by the dynamical map of $f^{(m)}_l$ onto  
\begin{equation}
\theta^{(m)}_l(m_T,p_T) = \frac{\tilde S^{(m)}_l(m_T,p_T)}{S_0(m_T,p_T)}\, .
\label{eq:thetaDefinition1}
\end{equation}
Due to the non-linear structure, the event-averaged spectrum is now in general affected by fluctuations,
\begin{equation}
\left( \frac{dN}{p_T dp_T d\phi d\eta_P} \right)_\text{event average} = S_0(m_T,p_T) \int D\tilde w\; p[\tilde w]\;\text{exp}\left[\sum_{m=-\infty}^\infty \sum_{l=1}^\infty \,\tilde w^{(m)}_l \, e^{i m \phi}\, \theta^{(m)}_l(m_T,p_T)\right].
\label{eq:particlespectrumaveragelog}
\end{equation}
For the particular case of a Gaussian probability distribution $p\left[ \tilde w \right]$ as 
discussed in Ref.~\cite{Floerchinger:2013vua}, one can perform the averaging in \eqref{eq:particlespectrumaveragelog} and one finds 
\begin{equation}
\begin{split}
S(m_T, p_T) = & \left( \frac{dN}{p_T dp_T d\phi d\eta_P} \right)_\text{event average} \\
= &  S_0(m_T,p_T) \times \exp\left[ \sum_{l=1}^\infty \langle \tilde w_l^{(0)} \rangle \theta^{(0)}_l(m_T,p_T) \right]\\
& \times \exp \left[ \frac{1}{2} \sum_{l_1,l_2=1}^\infty \theta^{(0)}_{l_1}(m_T,p_T)  \theta^{(0)}_{l_2}(m_T,p_T) \left( \langle \tilde w^{(0)}_{l_1} \tilde w^{(0)}_{l_2} \rangle - \langle \tilde w^{(0)}_{l_1} \rangle \langle \tilde w^{(0)}_{l_2} \rangle \right) \right]\\
& \times \exp \left[ \sum_{m=1}^\infty \sum_{l_1,l_2=1}^\infty \theta^{(m)}_{l_1}(m_T,p_T) \theta^{(m)}_{l_2}(m_T,p_T) \; \langle \tilde w^{(m)}_{l_1} \tilde w^{(m)*}_{l_2} \rangle  \right].
\end{split}
\label{eq:fullspectrum}
\end{equation}
Taking into account that $\langle \tilde w_l^{(m)} \rangle \propto \delta_{m0}$ for an 
azimuthally symmetric background, one sees easily that the spectra (\ref{eq:fullspectrum})
and (\ref{linearinS}) agree up to linear order in the fluctuations $\tilde w_l^{(m)}$. To quadratic
order in $\tilde w_l^{(m)}$, neither (\ref{eq:fullspectrum}) nor (\ref{linearinS}) is complete. 
However, since linear fluctuations in the fluid fields enter essentially in an exponential way,
we expect that (\ref{eq:fullspectrum}) resums relevant contributions to higher order in
the fluctuating fields. Therefore, we typically calculate the single inclusive hadron spectrum
from \eqref{eq:fullspectrum}.

\begin{figure}
\centering
\includegraphics[width=0.4\linewidth]{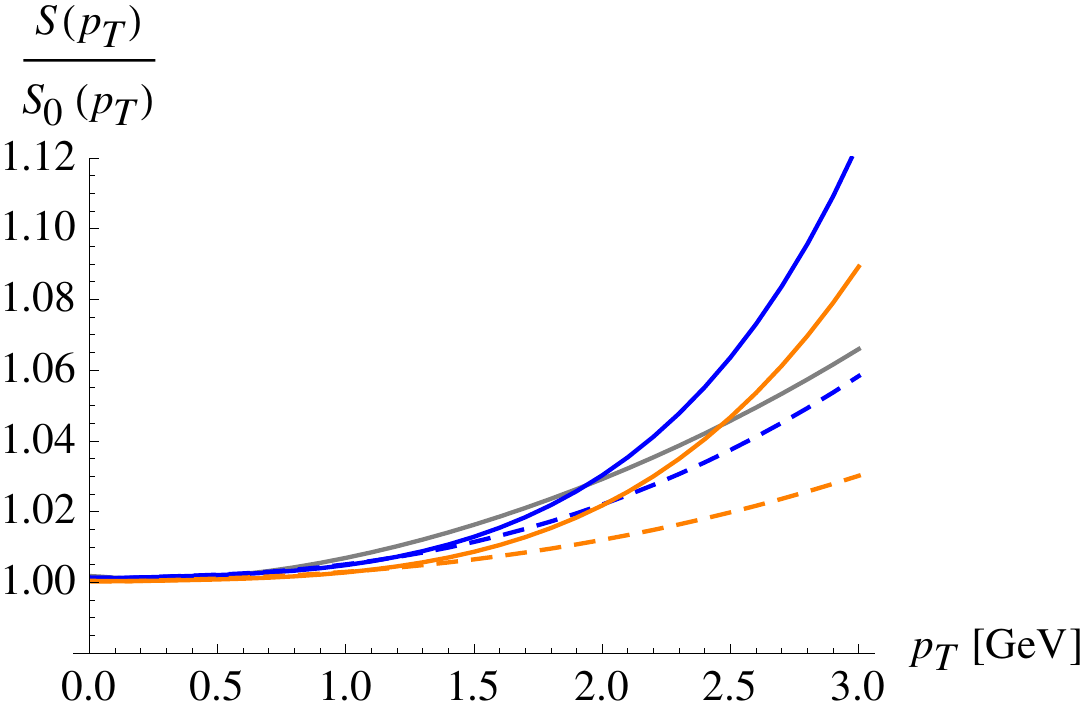}
\caption{Ratio of the full one-particle spectrum $S(p_T)$ to the background contribution $S_0(p_T)$. The deviation from 1 is due to event-by-event fluctuations (see \eqref{eq:fullspectrum}). The curves are for pion spectra calculated from ideal hydrodynamics, and viscous hydrodynamics with $\eta/s=0.08$ and $\eta/s=0.3$, respectively. The color coding is as in Fig.\ \ref{fig:FreezeOutCurve}. The dashed lines correspond to the result obtained when the shear viscous contributions at freeze-out (``quadratic ansatz'' in (\ref{eq3.1}) ) are neglected.}
\label{fig:fluctuationcontributionS}
\end{figure}

To quantify the effect of event-by-event fluctuations on the one-particle spectrum, 
we plot in Fig.\ \ref{fig:fluctuationcontributionS}
the ratio $S(m_T,p_T)/S_0(m_T,p_T)$ as defined by \eqref{eq:fullspectrum} 
for an ensemble of central collisions with initial conditions taken from a Glauber Monte-Carlo model as discussed in ref.\ \cite{Floerchinger:2013rya}. The deviation from 1 is relatively small, or, in other words, the one-particle spectrum is given to good approximation by the corresponding background part with only a small contribution coming from fluctuation effects. This figure \ref{fig:fluctuationcontributionS} can also be viewed as indicating that the difference between
the spectrum (\ref{linearinS}) (that reduces to $S_0(m_T,p_T)$ for central collisions 
with $\langle \tilde w_l^{(m)} \rangle =0$) and the spectrum 
$S(m_T,p_T)$ defined in \eqref{eq:fullspectrum} is relatively small. We believe that this observation holds also for other models of initial state fluctuations.

\subsection{Two-particle correlations and flow coefficients}
\label{sec4b}
Let us now discuss the two-particle spectrum. We use the standard definition for the ratio of the two particle spectrum (with both particles from the same event) to the product of two independently averaged one-particle spectra,
\begin{equation}
C_2(m_{T1},p_{T1},\phi_1;m_{T2},p_{T2},\phi_2) = \frac{\left(\frac{dN^\text{pairs}}{p_{T1} dp_{T1} d\phi_1 d\eta_P p_{T2} dp_{T2} d\phi_2 d\eta_P}\right)_\text{event average}}{\left(\frac{dN}{p_{T1}dp_{T1}d\phi_1 d\eta_P}\right)_\text{event average}\left(\frac{dN}{p_{T2}dp_{T2}d\phi_2 d\eta_P}\right)_\text{event average}}.
\end{equation}
Using an expansion similar to \eqref{eq:particlespectrumaveragelog} for both the numerator and denominator one finds
\begin{equation}
\begin{split}
C_2 = &  \int D\tilde w\; p[\tilde w]\;\text{exp}\left[\sum_{m=-\infty}^\infty \sum_{l=1}^\infty \,\tilde w^{(m)}_l \, \left[e^{i m \phi_1}\, \theta^{(m)}_l(m_{T1},p_{T1}) + e^{i m \phi_2}\, \theta^{(m)}_l(m_{T2},p_{T2})\right] \right] \\
& \times \left( \int D\tilde w\; p[\tilde w]\;\text{exp}\left[\sum_{m=-\infty}^\infty \sum_{l=1}^\infty \,\tilde w^{(m)}_l \, e^{i m \phi_1}\, \theta^{(m)}_l(m_{T1},p_{T1})\right] \right)^{-1}\\
& \times \left( \int D\tilde w\; p[\tilde w]\;\text{exp}\left[\sum_{m=-\infty}^\infty \sum_{l=1}^\infty \,\tilde w^{(m)}_l \, e^{i m \phi_2}\, \theta^{(m)}_l(m_{T2},p_{T2})\right] \right)^{-1}.
\end{split}
\end{equation}
For a Gaussian probability distribution $p[\tilde w]$, one can actually perform the functional integrals and one finds
\begin{equation}
\begin{split}
C_2 & = \exp \left[ \sum_{l_1,l_2=1}^\infty \theta^{(0)}_{l_1}(m_{T1},p_{T1}) \theta^{(0)}_{l_2}(m_{T2},p_{T2}) \left( \langle \tilde w^{(0)}_{l_1} \tilde w^{(0)}_{l_2} \rangle - \langle \tilde w^{(0)}_{l_1} \rangle \langle \tilde w^{(0)}_{l_2} \rangle \right)  \right]\\
& \times \exp \left[ \sum_{m=1}^\infty 2 \cos\left( m(\phi_1-\phi_2) \right)\sum_{l_1,l_2=1}^\infty \theta^{(m)}_{l_1}(m_{T1},p_{T1})  \theta^{(m)}_{l_2}(m_{T2},p_{T2}) \langle \tilde w^{(m)}_{l_1} \tilde w^{(m)*}_{l_2} \rangle \right].
\end{split}
\label{eq:twoparticlespectrumGaussian}
\end{equation}
The double differential harmonic flow coefficients $v_m\{2\}(m_{T1},p_{T1};m_{T2},p_{T2})$ are now defined by the expansion
\begin{equation}
\begin{split}
C_2(m_{T1},p_{T1},\phi_1;m_{T2},p_{T2},\phi_2) = &\;\; v_0^2\{2\}(m_{T1},p_{T1};m_{T2},p_{T2})\\
& +\sum_{m=1}^\infty 2 \cos\left(m(\phi_1-\phi_2)\right) \; v^2_m\{2\}(m_{T1},p_{T1};m_{T2},p_{T2}).
\end{split}
\label{eq:twoparticlevmexpansion}
\end{equation}
We note that the experimental determination of $C_2$ is usually made such that $v_0^2\{2\}(m_{T1},p_{T1}; m_{T2},p_{T2})$ or a $p_T$-integrated variant thereof, is normalized to $1$. In contrast, in our definition $v_0^2\{2\}$ does not necessarily equal $1$ even when integrated over $p_T$. This reflects a dispersion in event-by-event values of multiplicities, more specific the $p_T$-integrated $v_0^2\{2\}$ equals the ratio of $\langle N^2 \rangle$ and $\langle N \rangle^2$. We could adapt our normalization of $C_2$ to the experimental one by dropping the factor in the first line of \eqref{eq:twoparticlespectrumGaussian}. Since the argument of the exponential is typically small this leads to a small correction only.

In general, for very large fluctuations, the relation between \eqref{eq:twoparticlespectrumGaussian} and \eqref{eq:twoparticlevmexpansion} is non-trivial and the harmonic flow coefficients have to be calculated numerically by doing the appropriate Fourier transform. For not too large values of $v_m^2\{2\}$ one can expand the exponential in \eqref{eq:twoparticlespectrumGaussian} which yields
\begin{equation}
v_0^2\{2\}(m_{T1},p_{T1};m_{T2},p_{T2}) = 1 + \sum_{l_1,l_2=1}^\infty \theta^{(0)}_{l_1}(m_{T1},p_{T1}) \theta^{(0)}_{l_2}(m_{T2},p_{T2}) \left( \langle \tilde w^{(0)}_{l_1} \tilde w^{(0)}_{l_2} \rangle - \langle \tilde w^{(0)}_{l_1} \rangle \langle \tilde w^{(0)}_{l_2} \rangle \right)
\label{eq:v0doublelinear}
\end{equation}
for $m=0$ and
\begin{equation}
v_m^2\{2\}(m_{T1},p_{T1};m_{T2},p_{T2}) = \sum_{l_1,l_2=1}^\infty \theta^{(m)}_{l_1}(m_{T1},p_{T1})  \theta^{(m)}_{l_2}(m_{T2},p_{T2}) \langle \tilde w^{(m)}_{l_1} \tilde w^{(m)*}_{l_2} \rangle
\label{eq:vmdoublelinear}
\end{equation}
for $m\geq 0$, respectively. 

The $p_T$-integrated harmonic flow coefficients can be obtained by integrating the expression \eqref{eq:vmdoublelinear} over $p_{T1}$ and $p_{T2}$, appropriately weighted by the one-particle spectrum. More specific,
\begin{equation}
v_m^2\{2\} = \frac{\int dp_{T1} \int dp_{T2} \; p_{T1}\, p_{T2}\, S(m_{T1}, p_{T1}) S(m_{T2}, p_{T2}) \; v_m^2\{2\}(m_{T1},p_{T1} ; m_{T2},p_{T2}) }{ \left( \int dp_T \;p_T\, S(m_T,p_T) \right)^2}.
\end{equation}
The single-differential harmonic flow coefficients can be defined as
\begin{equation}
v_m\{2\}(m_T,p_T) = \frac{1}{v_m\{2\}} \frac{\int d\,p_{T2} \; p_{T2}\, S(m_{T2},p_{T2}) \;v_m^2\{2\}(m_{T1},p_{T1};m_{T2},p_{T2})}{\int dp_T\, p_T\, S(m_T,p_T)}
\end{equation}
and by construction they are normalized such that
\begin{equation}
v_m\{2\} = \frac{\int dp_T\; p_T\; S(m_T,p_T)\; v_m\{2\}(m_T,p_T)}{\int dp_T\; p_T\; S(m_T,p_T)}.
\end{equation}

We note that in general, the flow coefficients $v^2_m\{2\}(m_{T1},p_{T1};m_{T2},p_{T2})$
that enter the two-particle spectrum \eqref{eq:twoparticlevmexpansion} do not 
factorize into products $v_m\{2\}(m_{T1},p_{T1})\, v_m\{2\}(m_{T2},p_{T2})$. 
Therefore, testing the breakdown of such a factorization experimentally as in 
Ref.\cite{Aamodt:2011by} should not be 
interpreted in general as indicating the limited validity of a fluid dynamic description.

We note that Eqns.\ \eqref{eq:twoparticlespectrumGaussian}, \eqref{eq:v0doublelinear} and \eqref{eq:vmdoublelinear} constitute rather simple relations between the initial correlations in the expansion coefficients $w^{(m)}_l$ and the observable two-particle spectrum. The relation is given by the functions $\theta^{(m)}_l$ which can be calculated and tabulated in an appropriate way. 

\subsection{Generalizations}
\label{sec4c}
So far, we have concentrated our discussion on initial density fluctuations and vanishing rapidity dependence (Bjorken boost invariance). In general, one might expect that a detailed model of the initial state and early non-equilibrium dynamics predicts also the size and shape of initial fluctuations in other hydrodynamic fields, in particular fluid velocity and shear. Moreover, these fluctuations may have non-trivial rapidity dependence. We now generalize the most important expressions for the particle spectrum and harmonic flow coefficients to this case.

We chose the hydrodynamical background fields (\ref{eq3.17}) such that
the event-average over the fluctuating fields (\ref{eq3.18}) vanish, $\langle \tilde h_i \rangle=0$. The expansion of the hydrodynamical fields at the initial time $\tau_0$, Eq.\ \eqref{eq:FourierBesselEnthalpy} can be generalized to
\begin{equation}
h_i(\tau_0, r, \varphi, \eta) = \int \frac{d k_\eta}{2\pi} \sum_{m=-\infty}^\infty h^{(m)}_{i,l}(k_\eta) \; e^{im\varphi + i k_\eta \eta}\, f_{i,l}^{(m)}(r)\, ,
\label{eq:hinitialGeneral}
\end{equation}
where $f_{i,l}^{(m)}(r)$ are appropriate basis functions, see for example \cite{Floerchinger:2013vua}. Equation \eqref{eq:hinitialGeneral} describes the most general fluctuation around the azimuthal rotation and Bjorken boost symmetric background in the hydrodynamical fields at the initial time $\tau_0$. 

The contribution of each individual mode to the particle spectra at freeze-out can be determined in complete analogy to the density modes discussed above. The functions $\theta^{(m)}_l$ defined in \eqref{eq:thetaDefinition1} receive an additional index $i$ as well as a dependence on $k_\eta$ and become
\begin{equation}
\theta^{(m)}_{i, l}(k_\eta; m_T, p_T).
\end{equation}
Equation \eqref{eq:fullspectrum} which gives the one-particle spectrum for a Gaussian probability distribution of initial fluctuations gets generalized to
\begin{equation}
\begin{split}
& S(m_T, p_T) =  \left( \frac{dN}{p_T dp_T d\phi d\eta_P} \right)_\text{event average}\\
 &  = S_0(m_T,p_T) \times \exp {\Bigg [} \frac{1}{2} \sum_{m=-\infty}^\infty \int\frac{dk_\eta}{2\pi} \sum_{i_1,i_2,l_1,l_2}  \theta^{(m)}_{i_1, l_1}(k_\eta; m_T,p_T)\\ 
&  \times \theta^{(m)}_{i_2,l_2}(k_\eta; m_T,p_T) \; \langle h^{(m)}_{i_1, l_1}(k_\eta) h^{(m)*}_{i_2, l_2}(k_\eta) \rangle {\Bigg ]}.
\end{split}
\label{eq:fullspectrum2}
\end{equation}
Note that this does not depend on $\phi$ or $\eta_p$ in agreement with the assumed statistical symmetries. In a similar way, Eq.\ \eqref{eq:twoparticlespectrumGaussian} for the two-particle correlation function generalizes for $\langle \tilde h_i \rangle=0$ to
\begin{equation}
\begin{split}
C_2 & = \exp {\Bigg [} \sum_{m=-\infty}^\infty \int \frac{dk_\eta}{2\pi} \;\;e^{i m(\phi_1-\phi_2)}\; e^{i k_\eta(\eta_1-\eta_2)} \\
& \times \sum_{i_1,i_2,l_1,l_2} \theta^{(m)}_{i_1, l_1}(k_\eta;m_{T1},p_{T1})  \theta^{(m)}_{i_2, l_2}(k_\eta; m_{T2},p_{T2})^* \; \langle  h^{(m)}_{i_1, l_1}(k_\eta) h^{(m)}_{i_2, l_2}(k_\eta)^*\rangle {\Bigg ] }.
\end{split}
\label{eq:twoparticlespectrumGaussian2}
\end{equation}

\section{Conclusions}

We have discussed here the kinetic freeze-out for heavy ion collisions using a background-fluctuation splitting for the hydrodynamical fields. To this end, we have introduced in 
section~\ref{sec2} a version of the Cooper-Frye freeze-out prescription according to
which particles decouple instantaneously at constant background temperature rather
than at constant temperature. We have then discussed in section~\ref{sec3} how 
fluctuations in all fluid dynamic fields can be treated as perturbations on this
freeze-out hyper surface. In general, one can adopt a setting in which the
background solution is invariant with respect to azimuthal rotations and longitudinal 
boosts. The background contribution of the particle spectrum has these symmetries, as well. Deviations from these symmetries come then from event-by-event fluctuations in the hydrodynamical fields. In section~\ref{sec4}, we have decomposed arbitrary fluctuations 
in the initial conditions into an orthonormal set of basis functions.  To lowest (linear) order, these basis functions can be propagated  independently in hydrodynamics, and they can then be hadronized independently at freeze-out.  The central results of the present paper are \eqref{eq:fullspectrum} and \eqref{eq:twoparticlespectrumGaussian} that describe the single inclusive hadron spectrum and two-hadron correlations functions in terms of the hadronic response $\theta^{(m)}_{l}(m_{T},p_{T})$ of basis functions at freeze-out and the weights
that these basis functions carry in the initial conditions. After having calculated the 
hadronic response $\theta^{(m)}_{l}(m_{T},p_{T})$ for the basis functions only
one time, these equations allow one to determine the spectra and two-particle correlations
for arbitrary events without further fluid dynamic simulations. 

The formalism presented here is accurate to lowest (linear) order in fluctuating fields
only, and we did not attempt to go to quadratic and higher orders in the fluctuating hydrodynamical fields in this paper. We plan to study the numerical relevance 
of non-linear corrections in subsequent works, addressing the question of 
non-linear corrections in the fluid-dynamic evolution of basis functions, and the
question of additional non-linearities at freeze-out. Within the present paper, we
note only that the Cooper-Frye freeze-out formula \eqref{eq:CooperFrye} contains actually contributions from all orders in the velocity and temperature fields, since the
hydrodynamic fields enter in the exponent of occupation numbers as in Eq.\ \eqref{eq:occupationnumbersexpansion}. Therefore, although complete only to
leading order in fluctuations, the formalism presented here accounts at least for
some of the expected non-linear corrections. We also note that the azimuthal and boost invariance of the background that has greatly simplified our discussion to linear order 
will have important implications for higher order contributions, as well. For example, for a single event, the particle distribution may contain a mode $\sim e^{i m\phi}$ which at quadratic order has contributions from modes $e^{im_1 \varphi}$ and $e^{im_2 \varphi}$ in the hydrodynamic fields but these modes must fulfill $m_1+m_2=m$. The situation is similar with respect to Bjorken boost invariance: a mode $e^{ik_\eta \eta_P}$ may have contributions at quadratic order from two modes $e^{i k_{\eta 1}}$ and $e^{i k_{\eta 2}}$ but there is the constraint $k_\eta = k_{\eta 1} + k_{\eta 2}$.

Finally we emphasize that our current study is based on a sharp kinetic freeze-out and includes no phase of hadronic scatterings and resonance decays between the hydrodynamic regime and free streaming. In principle this can be incorporated into the mode-by-mode formalism by solving the corresponding kinetic equations for the background contribution and, in linearized form, for the fluctuations. In particular the influence of resonance decays can be quite sizable for certain observables, most prominent particle identified spectra and harmonic flow coefficients. The formula presented in the current paper could be used to initialize this hadronic scattering and decay phase although a description based on occupation numbers for interacting particles would be desirable.

\begin{appendix}
\section{Rapidity and azimuthal integrals}\label{sec:RapidtyAzimuthalIntegrals}
\label{appA}

In this appendix, we provide details on how to calculate from (\ref{eq3.1}) the
expressions for the spectrum of the background field  \eqref{eq:S0} and the 
contributions (\ref{eq:spectrumfluctkernels}) to first order in the fluctuating fields. 
For the background contribution, we calculate from equation (\ref{eq3.6})
the shear viscous term at freeze-out, 
\begin{equation}
\begin{split}
\frac{p_\mu p_\nu \pi_0^{\mu\nu}}{\epsilon_0 + p_0} = \tilde \pi_0^t {\bigg [} & m_T^2 (u_0^r)^2 \cosh^2(\eta_P-\eta) - 2 m_T p_T u_0^\tau u_0^r \cosh(\eta_P-\eta) \cos(\phi-\varphi) \\
& + p_T^2 (u_0^\tau)^2 \cos^2(\phi-\varphi) - p_T^2 \sin^2(\phi-\varphi) {\bigg ]} \\
+ \tilde \pi_0^{\eta\eta} {\bigg[ } & -\frac{1}{2}m_T^2 (u_0^r)^2 \cosh^2(\eta_P-\eta) + m_T p_T u_0^\tau u_0^r \cosh(\eta_P-\eta) \cos(\phi-\varphi) \\
& -\frac{1}{2} p_T^2 (u_0^\tau)^2 \cos^2(\phi-\varphi) - \frac{1}{2} p_T^2 \sin^2(\phi-\varphi) + m_T^2 \sinh^2(\eta_P-\eta) {\bigg ]}\, . 
\end{split}
\end{equation}
The integrals over space rapidity that appear in the Cooper-Frye freeze-out of different terms
of (\ref{eq3.1}) are then of the form
\begin{equation}
R_*(k,z) = \frac{1}{2}\int_{-\infty}^\infty d\eta \; e^{-z \cosh(\eta)} \; e^{ik\eta} 
f_*(\eta)\, ,
\end{equation}
where the integrand $f_*(\eta)$ on the right hand side is either unity (we write $*=0$)
or involves up to three powers in $\cosh\eta$ and $\sinh\eta$. We denote each such 
power by a subscript '$c$' or '$s$' respectively, so that for instance 
the integral $R_{ccc}$ is obtained for the integrand $f_*(\eta) = \cosh^3(\eta)$. 
All relevant integrals can then be expressed explicitly in terms of Bessel functions $K_n(z)$
of the second kind, 
\begin{align*}
& R_0(k,z)  = K_{i k}(z),\\
& R_c(k,z)  = \frac{1}{2} (K_{i k-1}(z)+K_{i k+1}(z)),\\
& R_s(k,z) = \frac{1}{2} (-K_{i k-1}(z)+K_{i k+1}(z)),\\
& R_{cc}(k,z) = \frac{1}{4} (K_{i k-2}(z)+2 K_{i
   k}(z)+K_{i k+2}(z)),\\
& R_{cs}(k,z) = \frac{1}{4} (-K_{i k-2}(z)+K_{i k+2}(z)), \tag{\stepcounter{equation}\theequation}\\
& R_{ss}(k,z)  = \frac{1}{4} (K_{i k-2}(z)-2 K_{i
   k}(z)+K_{i k+2}(z)),\\
& R_{ccc}(k,z) = \frac{1}{8} (K_{i k-3}(z)+3 K_{i k-1}(z)+3
   K_{i k+1}(z)+K_{i k+3}(z)),\\
& R_{ccs}(k,z) = \frac{1}{8} (-K_{i k-3}(z)-K_{i k-1}(z)+K_{i
   k+1}(z)+K_{i k+3}(z)),\\
& R_{css}(k,z)  = \frac{1}{8} (K_{i k-3}(z)-K_{i k-1}(z)-K_{i
   k+1}(z)+K_{i k+3}(z)).\\
\end{align*}
We define also the abbreviations
\begin{equation}
\begin{split}
& R_*(z)= R_*(0,z),
\end{split}
\end{equation}
to be used below.
Similarly one can perform the integrals over the azimuthal angle $\varphi$. Introducing
the shorthand 
\begin{equation}
A_*(m,z) = \frac{1}{2\pi}\int_{0}^{2\pi} d\varphi \; e^{z \cos(\varphi)} \; e^{im\varphi} 
g_*(\varphi)\, ,
\end{equation}
and considering for the integrand $g_*(\varphi)$ up to three powers in $\cos\varphi$ 
and $\sin\varphi$ such that for instance $A_{ccs}$ is the integral for 
$g_*(\varphi) = \cos^2\varphi\, \sin\varphi$, we find
\begin{align*}
& A_0(m,z) = I_{m}(z),\\
& A_c(m,z)  = \frac{1}{2} (I_{m-1}(z)+I_{m+1}(z)),\\
& A_s(m,z) = \frac{1}{2} (i I_{m-1}(z)-i I_{m+1}(z)),\\
& A_{cc}(m,z) = \frac{1}{4} (I_{m-2}(z)+2 I_m(z)+I_{m+2}(z)),\\
& A_{cs}(m,z)  = \frac{1}{4} (i I_{m-2}(z)-i I_{m+2}(z)),\tag{\stepcounter{equation}\theequation}\\
& A_{ss}(m,z) = \frac{1}{4} (-I_{m-2}(z)+2 I_m(z)-I_{m+2}(z)),\\
& A_{ccc}(m,z) = \frac{1}{8} (I_{m-3}(z)+3 I_{m-1}(z)+3
   I_{m+1}(z)+I_{m+3}(z)),\\
& A_{ccs}(m,z) = \frac{1}{8} (i I_{m-3}(z)+i I_{m-1}(z)-i
   I_{m+1}(z)-i I_{m+3}(z)),\\
& A_{css}(m,z) = \frac{1}{8}
   (-I_{m-3}(z)+I_{m-1}(z)+I_{m+1}(z)-I_{m+3}(z))\, .
\end{align*}
Below we will also use the abbreviation
\begin{equation}
A_*(z) = A_*(0,z)\, .
\end{equation}
The expressions above can be used to do the $\varphi$- and $\eta$-integrations of the
background contribution to the single inclusive hadron spectrum (\ref{eq3.1}). 
The final result Eq.\ \eqref{eq:S0} can then be expressed in terms of the integration kernels $Z_1$, $Z_2$, $Z_3$ and $Z_4$,  
\begin{align*}
Z_1(\tilde m_T, \tilde p_T, u_0^\tau, u_0^r,j)  = \frac{1+j}{2} & {\bigg [}  \tilde m_T^2 (u_0^r)^2 \;\; R_{ccc}(\tilde m_T u_0^\tau (1+j)) \;\; A_0(\tilde p_T u_0^r(1+j))\\*
&-2\tilde m_T \tilde p_T u_0^\tau u_0^r \;\; R_{cc}(\tilde m_T u_0^\tau (1+j)) \;\; A_{c}(\tilde p_T u_0^r(1+j))\\*
&+\tilde p_T^2 (u_0^\tau)^2 \;\; R_{c}(\tilde m_T u_0^\tau (1+j)) \;\; A_{cc}(\tilde p_T u_0^r(1+j))\\*
&-\tilde p_T^2 \;\;R_{c}(\tilde m_T u_0^\tau (1+j)) \;\; A_{ss}(\tilde p_T u_0^r(1+j)) {\bigg ]},\\
Z_2(\tilde m_T, \tilde p_T, u_0^\tau, u_0^r,j)  = \frac{1+j}{2} & {\bigg [}  \tilde m_T^2 (u_0^r)^2 \;\; R_{cc}(\tilde m_T u_0^\tau (1+j)) \;\; A_{c}(\tilde p_T u_0^r(1+j))\\*
&-2\tilde m_T \tilde p_T u_0^\tau u_0^r \;\; R_{c}(\tilde m_T u_0^\tau (1+j)) \;\; A_{cc}(\tilde p_T u_0^r(1+j))\\*
&+\tilde p_T^2 (u_0^\tau)^2 \;\; R_{0}(\tilde m_T u_0^\tau (1+j)) \;\; A_{ccc}(\tilde p_T u_0^r(1+j))\\*
&-\tilde p_T^2 \;\;R_{0}(\tilde m_T u_0^\tau (1+j)) \;\; A_{css}(\tilde p_T u_0^r(1+j)) {\bigg ]},\\
Z_3(\tilde m_T, \tilde p_T, u_0^\tau, u_0^r,j)  = \frac{1+j}{2} & {\bigg [}  -\frac{1}{2}\tilde m_T^2 (u_0^r)^2 \;\; R_{ccc}(\tilde m_T u_0^\tau (1+j)) \;\; A_0(\tilde p_T u_0^r(1+j))\\*
&+\tilde m_T \tilde p_T u_0^\tau u_0^r \;\; R_{cc}(\tilde m_T u_0^\tau (1+j)) \;\; A_{c}(\tilde p_T u_0^r(1+j))\\*
&-\frac{1}{2}\tilde p_T^2 (u_0^\tau)^2 \;\; R_{c}(\tilde m_T u_0^\tau (1+j)) \;\; A_{cc}(\tilde p_T u_0^r(1+j))\tag{\stepcounter{equation}\theequation}\\*
&-\frac{1}{2}\tilde p_T^2 \;\;R_{c}(\tilde m_T u_0^\tau (1+j)) \;\; A_{ss}(\tilde p_T u_0^r(1+j))\\*
& + \tilde m_T^2 \;\;R_{css}(\tilde m_T u_0^\tau (1+j)) \;\;A_0(\tilde p_T u_0^r(1+j)) {\bigg ]},\\
Z_4(\tilde m_T, \tilde p_T, u_0^\tau, u_0^r,j)  = \frac{1+j}{2} & {\bigg [}  -\frac{1}{2}\tilde m_T^2 (u_0^r)^2 \;\; R_{cc}(\tilde m_T u_0^\tau (1+j)) \;\; A_c(\tilde p_T u_0^r(1+j))\\*
&+\tilde m_T \tilde p_T u_0^\tau u_0^r \;\; R_{c}(\tilde m_T u_0^\tau (1+j)) \;\; A_{cc}(\tilde p_T u_0^r(1+j))\\*
&-\frac{1}{2}\tilde p_T^2 (u_0^\tau)^2 \;\; R_{0}(\tilde m_T u_0^\tau (1+j)) \;\; A_{ccc}(\tilde p_T u_0^r(1+j))\\*
&-\frac{1}{2}\tilde p_T^2 \;\;R_{0}(\tilde m_T u_0^\tau (1+j)) \;\; A_{css}(\tilde p_T u_0^r(1+j))\\*
& + \tilde m_T^2 \;\;R_{ss}(\tilde m_T u_0^\tau (1+j)) \;\;A_c(\tilde p_T u_0^r(1+j)) {\bigg ]}.
\end{align*}

To determine the contribution of fluctuation fields to the particle spectra we use the following integral kernels which are functions of the azimuthal wavenumber $m$, the rapidity wavenumber $k$, the transverse mass $\tilde m_T=m_T/T$, transverse momentum $\tilde p_T=p_T/T$, the chemical potential $\tilde \mu = \mu/T$, the temporal and radial components of the background fluid velocity $u_0^\tau$ and $u_0^r$, respectively, as well as the two independent components of the background shear stress tensor $\tilde \pi_0^t$ and $\tilde \pi_0^{\eta\eta}$ and the integer $j$,
\begin{equation}
Y_{\dots} = Y_{\dots}(m,k;\tilde m_T, \tilde p_T, \tilde\mu ; u_0^\tau, u_0^r, \tilde\pi_0^t, \tilde\pi_0^{\eta\eta};j).
\end{equation}
The functions $Y_{1a}$ and $Y_{1b}$ are also linear in the thermodynamic quantity $d \ln(T_0) / d \ln(w_0)$.
Explicit expressions are given by
\begin{align*}
Y_{1a}     = & \frac{d \ln (T_0)}{d \ln (w_0)} {\bigg [}\;
\tilde m_T u_0^\tau \;\; R_{cc}(k,\tilde m_T u_0^\tau (1+j)) \;\; A_{0}(m,\tilde p_T u_0^r (1+j))\\*
& - \tilde p_T u_0^r \;\;R_{c}(k,\tilde m_T u_0^\tau (1+j)) \;\; A_{c}(m,\tilde p_T u_0^r (1+j))\\*
& - \tilde \mu R_{c}(k,\tilde m_T u_0^\tau (1+j)) \;\; A_{0}(m,\tilde p_T u_0^r (1+j)),{\bigg ]}\\
Y_{1b}  = & \frac{d \ln (T_0)}{d \ln (w_0)}{\bigg [}\;
\tilde m_T u_0^\tau \;\; R_{c}(k,\tilde m_T u_0^\tau (1+j)) \;\; A_{c}(m,\tilde p_T u_0^r (1+j))\\*
& - \tilde p_T u_0^r \;\;R_{0}(k,\tilde m_T u_0^\tau (1+j)) \;\; A_{cc}(m,\tilde p_T u_0^r (1+j))\\*
& - \tilde \mu R_{0}(k,\tilde m_T u_0^\tau (1+j)) \;\; A_{c}(m,\tilde p_T u_0^r (1+j)) {\bigg ]},\\
%
% no-viscous contribution:
Y_{2a}  = & -\tilde m_T \tfrac{u_0^r}{u_0^\tau\sqrt{2}} \;\; R_{cc}(k,\tilde m_T u_0^\tau (1+j)) \;\; A_{0}(m,\tilde p_T u_0^r (1+j))\\*
& + \tilde p_T \tfrac{1}{\sqrt{2}} \;\; R_{c}(k,\tilde m_T u_0^\tau (1+j)) \;\; A_{0}(m-1,\tilde p_T u_0^r (1+j))\\*
% viscous contribution:
& + \tilde m_T^2 \tfrac{1}{\sqrt{2}} u_0^r \left( \tilde \pi_0^t - \tfrac{1}{2} \tilde \pi_0^{\eta\eta} \right)  \;\;R_{ccc}(k,\tilde m_T u_0^\tau (1+j)) \;\; A_0(m,\tilde p_T u_0^r (1+j))      \\*
& - \tilde m_T \tilde p_T \tfrac{1}{\sqrt{2}} \left(u_0^\tau+\tfrac{(u_0^r)^2}{u_0^\tau}\right) \left( \tilde \pi_0^t - \tfrac{1}{2} \tilde \pi_0^{\eta\eta} \right) \;\;R_{cc}(k,\tilde m_T u_0^\tau (1+j)) \;\; A_{c}(m,\tilde p_T u_0^r (1+j))     \\*
& + \tilde p_T^2 \tfrac{1}{\sqrt{2}} u_0^r \left( \tilde \pi_0^t - \tfrac{1}{2} \tilde \pi_0^{\eta\eta} \right)  \;\;R_{c}(k,\tilde m_T u_0^\tau (1+j)) \;\; A_{cc}(m,\tilde p_T u_0^r (1+j))      \\*
& - i\, \tilde m_T \tilde p_T \tfrac{1}{\sqrt{2}} \frac{1}{u_0^\tau} \left( \tilde \pi_0^t + \tfrac{1}{2} \tilde \pi_0^{\eta\eta} \right)  \;\;R_{cc}(k,\tilde m_T u_0^\tau (1+j)) \;\; A_{s}(m,\tilde p_T u_0^r (1+j)),      \\
%
% no-viscous contribution:
Y_{2b}  = & -\tilde m_T \tfrac{u_0^r}{u_0^\tau\sqrt{2}} \;\; R_{c}(k,\tilde m_T u_0^\tau (1+j)) \;\; A_{c}(m,\tilde p_T u_0^r (1+j))\\*
& + \tilde p_T \tfrac{1}{\sqrt{2}} \;\; R_{0}(k,\tilde m_T u_0^\tau (1+j)) \;\; A_{c}(m-1,\tilde p_T u_0^r (1+j)),\\*
% viscous contribution:
& + \tilde m_T^2 \tfrac{1}{\sqrt{2}} u_0^r \left( \tilde \pi_0^t - \tfrac{1}{2} \tilde \pi_0^{\eta\eta} \right)  \;\;R_{cc}(k,\tilde m_T u_0^\tau (1+j)) \;\; A_{c}(m,\tilde p_T u_0^r (1+j))      \\*
& - \tilde m_T \tilde p_T \tfrac{1}{\sqrt{2}} \left(u_0^\tau+\tfrac{(u_0^r)^2}{u_0^\tau}\right) \left( \tilde \pi_0^t - \tfrac{1}{2} \tilde \pi_0^{\eta\eta} \right) \;\;R_{c}(k,\tilde m_T u_0^\tau (1+j)) \;\; A_{cc}(m,\tilde p_T u_0^r (1+j))      \\*
& + \tilde p_T^2 \tfrac{1}{\sqrt{2}} u_0^r \left( \tilde \pi_0^t - \tfrac{1}{2} \tilde \pi_0^{\eta\eta} \right)  \;\;R_{0}(k,\tilde m_T u_0^\tau (1+j)) \;\; A_{ccc}(m,\tilde p_T u_0^r (1+j))      \\*
& - i\, \tilde m_T \tilde p_T \tfrac{1}{\sqrt{2}} \frac{1}{u_0^\tau} \left( \tilde \pi_0^t + \tfrac{1}{2} \tilde \pi_0^{\eta\eta} \right) \;\;R_{c}(k,\tilde m_T u_0^\tau (1+j))\;\; A_{cs}(m,\tilde p_T u_0^r (1+j)),      \\
%
% no-viscous:
Y_{3a}  = & -\tilde m_T \tfrac{u_0^r}{u_0^\tau\sqrt{2}} \;\; R_{cc}(k,\tilde m_T u_0^\tau (1+j)) \;\; A_{0}(m,\tilde p_T u_0^r (1+j)) \tag{\stepcounter{equation}\theequation} \\*
& + \tilde p_T \tfrac{1}{\sqrt{2}} \;\; R_{c}(k,\tilde m_T u_0^\tau (1+j)) \;\; A_{0}(m+1,\tilde p_T u_0^r (1+j)),\\*
% viscous:
& + \tilde m_T^2 \tfrac{1}{\sqrt{2}} u_0^r \left( \tilde \pi_0^t - \tfrac{1}{2} \tilde \pi_0^{\eta\eta} \right)  \;\;R_{ccc}(k,\tilde m_T u_0^\tau (1+j))\;\; A_0(m,\tilde p_T u_0^r (1+j))      \\*
& - \tilde m_T \tilde p_T \tfrac{1}{\sqrt{2}} \left(u_0^\tau+\tfrac{(u_0^r)^2}{u_0^\tau}\right) \left( \tilde \pi_0^t - \tfrac{1}{2} \tilde \pi_0^{\eta\eta} \right) \;\;R_{cc}(k,\tilde m_T u_0^\tau (1+j))\;\; A_{c}(m,\tilde p_T u_0^r (1+j))      \\*
& + \tilde p_T^2 \tfrac{1}{\sqrt{2}} u_0^r \left( \tilde \pi_0^t - \tfrac{1}{2} \tilde \pi_0^{\eta\eta} \right)  \;\;R_{c}(k,\tilde m_T u_0^\tau (1+j))\;\; A_{cc}(m,\tilde p_T u_0^r (1+j))      \\*
& + i\, \tilde m_T \tilde p_T \tfrac{1}{\sqrt{2}} \frac{1}{u_0^\tau} \left( \tilde \pi_0^t + \tfrac{1}{2} \tilde \pi_0^{\eta\eta} \right)  \;\;R_{cc}(k,\tilde m_T u_0^\tau (1+j))\;\; A_{s}(m,\tilde p_T u_0^r (1+j)),      \\
%
% viscous:
Y_{3b}  = & -\tilde m_T \tfrac{u_0^r}{u_0^\tau\sqrt{2}} \;\; R_{c}(k,\tilde m_T u_0^\tau (1+j)) \;\; A_{c}(m,\tilde p_T u_0^r (1+j))\\*
& + \tilde p_T \tfrac{1}{\sqrt{2}} \;\; R_{0}(k,\tilde m_T u_0^\tau (1+j)) \;\; A_{c}(m+1,\tilde p_T u_0^r (1+j)),\\*
% viscous contribution:
& + \tilde m_T^2 \tfrac{1}{\sqrt{2}} u_0^r \left( \tilde \pi_0^t - \tfrac{1}{2} \tilde \pi_0^{\eta\eta} \right)  \;\;R_{cc}(k,\tilde m_T u_0^\tau (1+j))\;\; A_{c}(m,\tilde p_T u_0^r (1+j))      \\*
& - \tilde m_T \tilde p_T \tfrac{1}{\sqrt{2}} \left(u_0^\tau+\tfrac{(u_0^r)^2}{u_0^\tau}\right) \left( \tilde \pi_0^t - \tfrac{1}{2} \tilde \pi_0^{\eta\eta} \right) \;\;R_{c}(k,\tilde m_T u_0^\tau (1+j))\;\; A_{cc}(m,\tilde p_T u_0^r (1+j))      \\*
& + \tilde p_T^2 \tfrac{1}{\sqrt{2}} u_0^r \left( \tilde \pi_0^t - \tfrac{1}{2} \tilde \pi_0^{\eta\eta} \right)  \;\;R_{0}(k,\tilde m_T u_0^\tau (1+j))\;\; A_{ccc}(m,\tilde p_T u_0^r (1+j))      \\*
& + i\, \tilde m_T \tilde p_T \tfrac{1}{\sqrt{2}} \frac{1}{u_0^\tau} \left( \tilde \pi_0^t + \tfrac{1}{2} \tilde \pi_0^{\eta\eta} \right)  \;\;R_{c}(k,\tilde m_T u_0^\tau (1+j))\;\; A_{cs}(m,\tilde p_T u_0^r (1+j)),      \\
%
% no-viscous:
Y_{4a}  = & -\tilde m_T \;\; R_{cs}(k,\tilde m_T u_0^\tau (1+j)) \;\; A_{0}(m,\tilde p_T u_0^r (1+j))\\*
% viscous:
& - \tilde m_T^2 \frac{1}{u_0^\tau} \pi_0^{\eta\eta} \;\;R_{ccs}(k,\tilde m_T u_0^\tau (1+j)) \;\; A_{0}(m,\tilde p_T u_0^r (1+j)),\\
%
% no-viscous:
Y_{4b}  = & -\tilde m_T \;\; R_{s}(k,\tilde m_T u_0^\tau (1+j)) \;\; A_{c}(m,\tilde p_T u_0^r (1+j))\\*
% viscous:
& - \tilde m_T^2 \frac{1}{u_0^\tau} \pi_0^{\eta\eta} \;\;R_{cs}(k,\tilde m_T u_0^\tau (1+j)) \;\; A_{c}(m,\tilde p_T u_0^r (1+j)),\\
\end{align*}

The integral kernels $Y_{5a}$ and $Y_{5b}$ are reserved for the contribution from bulk viscous terms. In the present work we neglect these. The kernels determining the contribution from fluctuations in the shear stress are
\begin{align*} 
Y_{6a}  = &\; i\, \tilde p_T^2 \tfrac{1}{2\sqrt{2}} \;\; R_{c}(k,\tilde m_T u_0^\tau (1+j)) \;\; A_{ss}(m,\tilde p_T u_0^r (1+j))\\*
& + \tilde p_T^2 \tfrac{1}{\sqrt{2}} \;\;R_{c}(k,\tilde m_T u_0^\tau (1+j)) \;\; A_{cs}(m,\tilde p_T u_0^r (1+j))\\*
& - i \tilde m_T^2 \tfrac{1}{2\sqrt{2}} (u_0^r)^2 \;\;R_{ccc}(k,\tilde m_T u_0^\tau (1+j)) \;\; A_{0}(m,\tilde p_T u_0^r (1+j))\\*
& + i \tilde m_T \tilde p_T \tfrac{1}{\sqrt{2}} u_0^\tau u_0^r \;\;R_{cc}(k,\tilde m_T u_0^\tau (1+j)) \;\; A_{c}(m,\tilde p_T u_0^r (1+j))\\*
& - i \tilde p_T^2 \tfrac{1}{2\sqrt{2}} (u_0^\tau)^2 \;\;R_{c}(k,\tilde m_T u_0^\tau (1+j)) \;\; A_{cc}(m,\tilde p_T u_0^r (1+j))\\*
& - \tilde m_T \tilde p_T \tfrac{1}{\sqrt{2}} \frac{u_0^r}{u_0^\tau} \;\;R_{cc}(k,\tilde m_T u_0^\tau (1+j)) \;\; A_{s}(m,\tilde p_T u_0^r (1+j)),\\
Y_{6b}  = &\; i\, \tilde p_T \tfrac{1}{2\sqrt{2}} \;\; R_{0}(k,\tilde m_T u_0^\tau (1+j)) \;\; A_{css}(m,\tilde p_T u_0^r (1+j))\\*
& + \tilde p_T^2 \tfrac{1}{\sqrt{2}} \;\;R_{0}(k,\tilde m_T u_0^\tau (1+j)) \;\; A_{ccs}(m,\tilde p_T u_0^r (1+j))\\*
& - i \tilde m_T^2 \tfrac{1}{2\sqrt{2}} (u_0^r)^2 \;\;R_{cc}(k,\tilde m_T u_0^\tau (1+j)) \;\; A_{c}(m,\tilde p_T u_0^r (1+j))\\*
& + i \tilde m_T \tilde p_T \tfrac{1}{\sqrt{2}} u_0^\tau u_0^r \;\;R_{c}(k,\tilde m_T u_0^\tau (1+j)) \;\; A_{cc}(m,\tilde p_T u_0^r (1+j))\\*
& - i \tilde p_T^2 \tfrac{1}{2\sqrt{2}} (u_0^\tau)^2 \;\;R_{0}(k,\tilde m_T u_0^\tau (1+j)) \;\; A_{ccc}(m,\tilde p_T u_0^r (1+j))\\*
& - \tilde m_T \tilde p_T \tfrac{1}{\sqrt{2}} \frac{u_0^r}{u_0^\tau} \;\;R_{c}(k,\tilde m_T u_0^\tau (1+j)) \;\; A_{cs}(m,\tilde p_T u_0^r (1+j)), \\
Y_{7a}  = &- i\, \tilde p_T^2 \tfrac{1}{2\sqrt{2}} \;\; R_{c}(k,\tilde m_T u_0^\tau (1+j)) \;\; A_{ss}(m,\tilde p_T u_0^r (1+j))\\*
& + \tilde p_T^2 \tfrac{1}{\sqrt{2}} \;\;R_{c}(k,\tilde m_T u_0^\tau (1+j)) \;\; A_{cs}(m,\tilde p_T u_0^r (1+j))\\*
& + i \tilde m_T^2 \tfrac{1}{2\sqrt{2}} (u_0^r)^2 \;\;R_{ccc}(k,\tilde m_T u_0^\tau (1+j)) \;\; A_{0}(m,\tilde p_T u_0^r (1+j))\\*
& - i \tilde m_T \tilde p_T \tfrac{1}{\sqrt{2}} u_0^\tau u_0^r \;\;R_{cc}(k,\tilde m_T u_0^\tau (1+j)) \;\; A_{c}(m,\tilde p_T u_0^r (1+j))\\*
& + i \tilde p_T^2 \tfrac{1}{2\sqrt{2}} (u_0^\tau)^2 \;\;R_{c}(k,\tilde m_T u_0^\tau (1+j)) \;\; A_{cc}(m,\tilde p_T u_0^r (1+j))\\*
& - \tilde m_T \tilde p_T \tfrac{1}{\sqrt{2}} \frac{u_0^r}{u_0^\tau} \;\;R_{cc}(k,\tilde m_T u_0^\tau (1+j)) \;\; A_{s}(m,\tilde p_T u_0^r (1+j)),\tag{\stepcounter{equation}\theequation} \\
Y_{7b}  = &- i\, \tilde p_T \tfrac{1}{2\sqrt{2}} \;\; R_{0}(k,\tilde m_T u_0^\tau (1+j)) \;\; A_{css}(m,\tilde p_T u_0^r (1+j))\\*
& + \tilde p_T^2 \tfrac{1}{\sqrt{2}} \;\;R_{0}(k,\tilde m_T u_0^\tau (1+j)) \;\; A_{ccs}(m,\tilde p_T u_0^r (1+j)),\\*
& + i \tilde m_T^2 \tfrac{1}{2\sqrt{2}} (u_0^r)^2 \;\;R_{cc}(k,\tilde m_T u_0^\tau (1+j)) \;\; A_{c}(m,\tilde p_T u_0^r (1+j))\\*
& - i \tilde m_T \tilde p_T \tfrac{1}{\sqrt{2}} u_0^\tau u_0^r \;\;R_{c}(k,\tilde m_T u_0^\tau (1+j)) \;\; A_{cc}(m,\tilde p_T u_0^r (1+j))\\*
& + i \tilde p_T^2 \tfrac{1}{2\sqrt{2}} (u_0^\tau)^2 \;\;R_{0}(k,\tilde m_T u_0^\tau (1+j)) \;\; A_{ccc}(m,\tilde p_T u_0^r (1+j))\\*
& - \tilde m_T \tilde p_T \tfrac{1}{\sqrt{2}} \frac{u_0^r}{u_0^\tau} \;\;R_{c}(k,\tilde m_T u_0^\tau (1+j)) \;\; A_{cs}(m,\tilde p_T u_0^r (1+j)), \\
Y_{8a}  = & \tilde m_T \tilde p_T \tfrac{1}{\sqrt{2}} \;\; R_{cs}(k,\tilde m_T u_0^\tau (1+j)) \;\; A_{c}(m,\tilde p_T u_0^r (1+j))\\*
& + i \tilde m_T \tilde p_T \tfrac{1}{\sqrt{2}} \;\;R_{cs}(k,\tilde m_T u_0^\tau (1+j)) \;\; A_{s}(m,\tilde p_T u_0^r (1+j))\\*
& - \tilde m_T^2 \tfrac{1}{\sqrt{2}} \tfrac{u_0^r}{u_0^\tau} \;\;R_{cc}(k,\tilde m_T u_0^\tau (1+j)) \;\; A_{s}(m,\tilde p_T u_0^r (1+j))\\
Y_{8b}  = & \tilde m_T \tilde p_T \tfrac{1}{\sqrt{2}} \;\; R_{s}(k,\tilde m_T u_0^\tau (1+j)) \;\; A_{cc}(m,\tilde p_T u_0^r (1+j))\\*
& + i \tilde m_T \tilde p_T \tfrac{1}{\sqrt{2}} \;\;R_{s}(k,\tilde m_T u_0^\tau (1+j)) \;\; A_{cs}(m,\tilde p_T u_0^r (1+j))\\*
& - \tilde m_T^2 \tfrac{1}{\sqrt{2}} \tfrac{u_0^r}{u_0^\tau} \;\;R_{c}(k,\tilde m_T u_0^\tau (1+j)) \;\; A_{cs}(m,\tilde p_T u_0^r (1+j)),\\
Y_{9a}  = & \tilde m_T \tilde p_T \tfrac{1}{\sqrt{2}} \;\; R_{cs}(k,\tilde m_T u_0^\tau (1+j)) \;\; A_{c}(m,\tilde p_T u_0^r (1+j))\\*
& - i \tilde m_T \tilde p_T \tfrac{1}{\sqrt{2}} \;\;R_{cs}(k,\tilde m_T u_0^\tau (1+j)) \;\; A_{s}(m,\tilde p_T u_0^r (1+j))\\*
& - \tilde m_T^2 \tfrac{1}{\sqrt{2}} \tfrac{u_0^r}{u_0^\tau} \;\;R_{cc}(k,\tilde m_T u_0^\tau (1+j)) \;\; A_{s}(m,\tilde p_T u_0^r (1+j)),\\
Y_{9b}  = & \tilde m_T \tilde p_T \tfrac{1}{\sqrt{2}} \;\; R_{s}(k,\tilde m_T u_0^\tau (1+j)) \;\; A_{cc}(m,\tilde p_T u_0^r (1+j))\\*
& - i \tilde m_T \tilde p_T \tfrac{1}{\sqrt{2}} \;\;R_{s}(k,\tilde m_T u_0^\tau (1+j)) \;\; A_{cs}(m,\tilde p_T u_0^r (1+j))\\*
& - \tilde m_T^2 \tfrac{1}{\sqrt{2}} \tfrac{u_0^r}{u_0^\tau} \;\;R_{c}(k,\tilde m_T u_0^\tau (1+j)) \;\; A_{cs}(m,\tilde p_T u_0^r (1+j)),\\
Y_{10a}  = & \tilde m_T^2 \tfrac{1}{2} \;\; R_{css}(k,\tilde m_T u_0^\tau (1+j)) \;\; A_{0}(m,\tilde p_T u_0^r (1+j))\\*
& - \tilde p_T^2 \tfrac{1}{4} \;\; R_{c}(k,\tilde m_T u_0^\tau (1+j)) \;\; A_{ss}(m,\tilde p_T u_0^r (1+j))\\*
& - \tilde m_T^2 \tfrac{1}{4} (u_0^r)^2 \;\; R_{ccc}(k,\tilde m_T u_0^\tau (1+j)) \;\; A_{0}(m,\tilde p_T u_0^r (1+j))\\*
& + \tilde m_T \tilde p_T \tfrac{1}{2} u_0^\tau u_0^r \;\; R_{cc}(k,\tilde m_T u_0^\tau (1+j)) \;\; A_{c}(m,\tilde p_T u_0^r (1+j))\\*
& - \tilde p_T^2 \tfrac{1}{4} (u_0^\tau)^2 \;\; R_{c}(k,\tilde m_T u_0^\tau (1+j)) \;\; A_{cc}(m,\tilde p_T u_0^r (1+j)),\\
Y_{10b}  = & \tilde m_T^2 \tfrac{1}{2} \;\; R_{ss}(k,\tilde m_T u_0^\tau (1+j)) \;\; A_{c}(m,\tilde p_T u_0^r (1+j))\\*
& - \tilde p_T^2 \tfrac{1}{4} \;\; R_{0}(k,\tilde m_T u_0^\tau (1+j)) \;\; A_{css}(m,\tilde p_T u_0^r (1+j))\\*
& - \tilde m_T^2 \tfrac{1}{4} (u_0^r)^2 \;\; R_{cc}(k,\tilde m_T u_0^\tau (1+j)) \;\; A_{c}(m,\tilde p_T u_0^r (1+j))\\*
& + \tilde m_T \tilde p_T \tfrac{1}{2} u_0^\tau u_0^r \;\; R_{c}(k,\tilde m_T u_0^\tau (1+j)) \;\; A_{cc}(m,\tilde p_T u_0^r (1+j))\\*
& - \tilde p_T^2 \tfrac{1}{4} (u_0^\tau)^2 \;\; R_{0}(k,\tilde m_T u_0^\tau (1+j)) \;\; A_{ccc}(m,\tilde p_T u_0^r (1+j)).
\end{align*}
With the help of the analytic expressions compiled in this appendix one can reduce the three-dimensional integration over the freeze-out surface to a one-dimensional integral along a curve in the $\tau$-$r$-plane that can be easily done numerically.

\section{Freeze-out at constant temperature versus freeze-out at constant background temperature}\label{sec:FreezeOutTemperature}

In this appendix we compare the freeze-out condition used throughout the paper, which is based on a kinetic freeze-out at constant background temperature, to the freeze-out at constant total temperature. We prove that to linear order in the perturbations of the fluid dynamic fields around the background, the difference between the two prescriptions vanishes.

We start from eq.\ \eqref{eq:CooperFrye}. For a freeze-out at constant background temperature the surface $\Sigma_f$ is independent of the perturbation in fluid fields $T_1(x)$, $u_1^\mu(x)$ etc.\ but the distribution function depends on them. In contrast, for a freeze-out at fixed total temperature, the position of the surface (which we denote as $\tilde \Sigma_{f}$) does depend on $T_1$. The $T_1$-dependence of the distribution function $f_i$ drops out.

The difference in between the two descriptions can be written as
\begin{equation}
\begin{split}
E \left( \frac{dN}{d^3 p}{\bigg |}_{T=\text{const}} - \frac{dN}{d^3 p}{\Bigg |}_{T_0=\text{const}} \right) = & - \frac{1}{(2\pi)^3} p_\mu \left[\int d \tilde \Sigma_f^\mu \; f_i (p^\mu; T_\text{fo}, u^\mu(x),\ldots) - \int d\Sigma_f^\mu \; f_i (p^\mu; T(x), u^\mu(x),\ldots) \right] \\
= & - \frac{1}{(2\pi)^3} p^\mu \int d^4 x \; \partial_\mu f_i(p^\mu; T(x), u^\mu(x), \ldots).
\end{split}
\end{equation}
In the last equation we have used Gauss' law and the integral in the last line goes over the four-dimensional volume between the two three-dimensional freeze-out surfaces $\tilde \Sigma_f$ and $\Sigma_f$. It is clear that this volume vanishes when $T_1 \to 0$ and in fact one can write to linear order in fluctuation fields
\begin{equation}
\begin{split}
E \left( \frac{dN}{d^3 p}{\bigg |}_{T=\text{const}} - \frac{dN}{d^3 p}{\Bigg |}_{T_0=\text{const}} \right) = &  -\frac{1}{(2\pi)^3} p^\mu \int d\tilde \Sigma_{f}^\nu \; \Delta x_\nu \; \partial_\mu f_i(p^\mu; T(x), u^\mu(x), \ldots),
\end{split}
\label{eq:B2}
\end{equation}
where $\Delta x^\nu$ is the difference between a point on the surface $\Sigma_f$ and one on $\tilde\Sigma_f$. Now one can use the expression for energy current within kinetic theory,
\begin{equation}
T^{\mu0}(x) = \sum_i \int \frac{d^3 p}{(2\pi)^3} \; p^\mu \, f_i(p^\mu; T(x), \ldots). 
\end{equation}
At kinetic freeze-out, by definition, this is conserved not only as a whole, $\partial_\mu T^{\mu 0}(x) = 0$, but actually for each particle species and momentum mode,
\begin{equation}
p^\mu \partial_\mu f_i(p^\mu; T(x), \ldots) = 0.
\end{equation}
Using this in eq.\ \eqref{eq:B2} shows that the difference between the freeze-out at constant temperature and the freeze-out at constant background temperature indeed vanishes to linear order in the fluctuating hydrodynamical fields. 

More generally, to higher orders in fluctuating fields, one can still write the particle spectrum as an integral over the freeze-out curve determined for the background solution although there might be small correction terms accounting for the fact that freeze-out happens at fixed fluctuating temperature. On the other side, in the absence of a precise theory where and how freeze-out really happens, we do not see a physics argument to prefer one of the prescriptions. Therefore, one may simply work with the freeze-out at strictly constant background temperature to all orders in fluctuations. This should be fine also if the hydrodynamic description is supplemented by an additional phase of hadronic scatterings and decays described by kinetic theory.

\end{appendix}

\end{document}